\documentclass[12pt]{article}

\usepackage[T5]{fontenc}
\usepackage[margin=1in]{geometry}
\usepackage[labelfont=bf]{caption}
\usepackage{tabularray}

\usepackage{algorithm,algpseudocode,amsmath,amssymb,amsthm,float,mathtools,thmtools,thm-restate,scalerel,stackengine,verbatim,subcaption,bbm}
\usepackage[shortlabels]{enumitem}
\usepackage{tikz}
\usepackage[hidelinks]{hyperref}
\usepackage[capitalize]{cleveref}

\makeatletter
\@ifundefined{newcounteralias}{}{%
  \renewcommand\thmt@autorefsetup{%
    \expandafter\def
      \csname\thmt@envname autorefname\expandafter\endcsname
      \expandafter{\thmt@thmname}}%
}
\makeatother

\usetikzlibrary{decorations.pathreplacing, positioning}
 
\stackMath
\newcommand\reallywidehat[1]{%
\savestack{\tmpbox}{\stretchto{%
  \scaleto{%
      \scalerel*[\widthof{\ensuremath{#1}}]{\kern-.6pt\bigwedge\kern-.6pt}%
          {\rule[-\textheight/2]{1ex}{\textheight}}
            }{\textheight}%
            }{0.5ex}}%
            \stackon[1pt]{#1}{\tmpbox}%
            }
\newcommand{\bool}{\set{0,1}}

\newcommand{\Cbb}{\mathbb{C}}
\newcommand{\Fbb}{\mathbb{F}}

\newcommand{\Sbb}{\mathbb{S}}
\newcommand{\Zbb}{\mathbb{Z}}
\newcommand{\Pbb}{\mathbb{P}}

\newcommand{\Pb}{\mathbf{P}}

\newcommand{\Ac}{\mathcal{A}}
\newcommand{\Bc}{\mathcal{B}}
\newcommand{\Cc}{\mathcal{C}}
\newcommand{\Dc}{\mathcal{D}}

\newcommand{\Mc}{\mathcal{M}}
\newcommand{\Pc}{\mathcal{P}}

\newcommand{\Sc}{\mathcal{S}}

\newcommand{\supp}{\text{\normalfont{supp}}}

\newcommand{\wt}[1]{\widetilde{#1}}
\newcommand{\wh}[1]{\widehat{#1}}
\newcommand{\eps}{\varepsilon}
\newcommand{\Tr}{\text{Tr}}

\newcommand{\prob}[2][]{\operatorname*{\mathbb{P}}_{#1 }\brac*{#2}}
\newcommand{\E}[2][]{\operatorname*{\mathbb{E}}_{#1 }\brac*{#2}}
\newcommand{\Var}[2][]{\operatorname*{\normalfont{\text{Var}}}_{#1 }\paren*{#2}}

\newcommand{\bO}[1]{\operatorname*{O}\paren*{#1}}
\newcommand{\bOt}[1]{\operatorname*{\wt{O}}\paren*{#1}}

\newcommand{\bOm}[1]{\operatorname*{\Omega}\paren*{#1}}

\newcommand{\bT}[1]{\operatorname*{\Theta}\paren*{#1}}
\newcommand{\bTt}[1]{\operatorname*{\wt{\Theta}}\paren*{#1}}

\newcommand{\plog}{\operatorname*{\normalfont{\text{polylog}}}}

\DeclareMathOperator*{\argmin}{arg\,min}

\DeclareMathOperator{\poly}{poly}

\DeclarePairedDelimiter\ceil{\lceil}{\rceil}
\DeclarePairedDelimiter\norm{\lVert}{\rVert}
\DeclarePairedDelimiter\abs{|}{|}
\DeclarePairedDelimiter\brac{\lbrack}{\rbrack}
\DeclarePairedDelimiter\set{\lbrace}{\rbrace}
\DeclarePairedDelimiter\paren{\lparen}{\rparen}

\DeclarePairedDelimiter\bra{\langle}{|}
\DeclarePairedDelimiter\ket{|}{\rangle}
\DeclarePairedDelimiterX\braket[2]{\langle}{\rangle}{#1\delimsize\vert#2}
\newcommand{\ketbra}[2]{\ket{#1}\!\bra{#2}}

\newtheorem{theorem}{Theorem}
\newtheorem{lemma}[theorem]{Lemma}
\newtheorem{fact}[theorem]{Fact}
\newtheorem{corollary}[theorem]{Corollary}

\theoremstyle{definition}
\newtheorem{definition}[theorem]{Definition}
\theoremstyle{remark}

\newcommand{\nth}{^\text{th}}

{\makeatletter
	\gdef\commentmark{%
		\expandafter\ifx\csname @mpargs\endcsname\relax 
		\expandafter\ifx\csname @captype\endcsname\relax 
		\marginpar{comment}
		\else
		comment 
		\fi
		\else
		comment 
		\fi}
	\gdef\comment{\@ifnextchar[\comment@lab\comment@nolab}
	\long\gdef\comment@lab[#1]#2{{\bf [\commentmark #2 ---{\sc #1}]}}
	\long\gdef\comment@nolab#1{{\bf [\commentmark #1]}}
}

\title{Hamiltonian Property Testing via Block Encoding}
\date{}
\author{Aprup Kale\\ National University of Singapore\\
\texttt{aprup.kale@nus.edu.sg} \and John Kallaugher\\ National University of
Singapore\\ \texttt{john.kallaugher@nus.edu.sg}}

\begin{document}
\maketitle
\begin{abstract}
\noindent
We give a method for constructing a block-encoding of an unknown Hamiltonian
$H$ with $\varepsilon$ error from its time evolution operator $e^{-iHt}$ in
polylogarithmic total evolution time. This immediately allows sampling from the
Pauli spectrum of $H$ with almost-optimal overhead
$\widetilde{\textnormal{O}}(\|H\|_2^{-2})$.

As a consequence we develop new tolerant Hamiltonian property testing
algorithms, which determine whether an unknown Hamiltonian is within
$\varepsilon_1$ of having a given property, or $\varepsilon_2$-far. We give
algorithms for testing locality, sparsity, degree-boundedness, being a junta,
and Pauli dimensionality, all with total evolution time scaling with
$1/(\varepsilon_2 - \varepsilon_1)^2$. We also apply our block-encoding
technique directly to obtain an $\widetilde{\textnormal{O}}(1/(\varepsilon_2 -
\varepsilon_1)^2)$-time tolerant algorithm for Hamiltonian certification.

Our algorithms require neither inverse nor controlled application of the time
evolution unitary, and in each case, our algorithm matches or improves on the
best non-tolerant testing algorithm. By a reduction to (Guti\'errez,~Lee,~and
Zur~`26), we show that all of these problems require at least
$\Omega(1/(\varepsilon_2(\varepsilon_2 - \varepsilon_1)))$ evolution time.
\end{abstract}

\section{Introduction}
The Hamiltonian of a quantum (or classical) system is an operator that
describes how that system evolves in the absence of external interactions. In
the standard computer science setting, we consider a system on $n$ qubits, and
the Hamiltonian is a $2^n$ by $2^n$ Hermitian matrix with an associated
\emph{time evolution operator} $e^{-iHt}$, where applying $e^{-iHt}$ to the
system corresponds to evolving it for time $t$.

Suppose the only way we can access an unknown Hamiltonian is by applying this
time evolution operator. What can we discover about $H$? This question has been
well-studied when it comes to learning $H$: finding a description of $H$ in its
entirety. Recent work, starting with~\cite{BCO24}, has considered the related
problem of property testing in this model. There is some set $\Pc$ of
Hamiltonians, and we want to know whether $H$ is in $\Pc$, or if it is
$\eps$-far from every Hamiltonian in $\Pc$. 

The standard distance model is the ``normalized Frobenius norm''
$2^{-n/2}\norm{H}_\text{F}$, which is also the (unnormalized) 2-norm of $H$ in
the Pauli basis $\set{I, X, Y, Z}^{\otimes n}$. In the ``tolerant'' variation
of this problem, we instead have two thresholds $\eps_2 > \eps_1 > 0$, and are
asked to distinguish between $H$ being within $\eps_1$ of some Hamiltonian in
$\Pc$, and $H$ being at least $\eps_2$ from every Hamiltonian in $\Pc$.  The
cost measure here is total evolution time: if the $j\nth$ application of the
time evolution operator is for $e^{-iHt_j}$, it costs $t_j$, and the total time
cost of the algorithm is $\sum_j t_j$.

As in the better-studied unitary\footnote{In the unitary property testing
setting, we typically have access to the inverse operator, and so these methods
are usually paired with a quantum amplitude estimation step that improves the
dependence on the error parameters by a quadratic factor. In the Hamiltonian
property testing setting we are usually \emph{not} allowed to set $t < 0$, and
therefore cannot access the inverse of the time evolution operator. It has been
shown~\cite{TW25,GLZ26,allen2026power} that this rules out a generic
application of QAE.} property testing setting, a ubiquitous tool in Hamiltonian
property testing is \emph{Pauli sampling}.

\subsection{Pauli Sampling}
Any $n$-qubit operator $A$ can be written in the (orthonormal) Pauli basis
$\Pbb_n = \set{I,X,Y,Z}^{\otimes n}$ as \[
  A = \sum_{P \in \Pbb_n} \alpha_P P.
\]
 When $A$ is a unitary, the squared absolute values of the
coefficients $(\abs{\alpha_P}^2)_{P \in \Pbb_n}$ form a probability distribution.
There is a simple procedure for sampling from this distribution with a single
application of $A$: prepare the ``Choi state'' \[
  \ket{A} \coloneqq 2^{-n/2}(I \otimes A)\sum_{i \in \bool^n} \ket{ii} = \sum_{P \in
  \Pbb_n} \alpha_P \ket{P}
\]
and then measure it in the orthonormal basis given by the Choi states of the
Paulis $\ket{P}$ (the orthonormality of the Paulis under the Hilbert-Schmidt
inner product guarantees orthonormality of this basis under vector inner
products). 

This procedure can be seen as a generalization of ``Fourier sampling'' for
Boolean functions and is one of the key tools in testing properties of
unitaries \cite{MdW16}. It is therefore unsurprising that most work on
Hamiltonian property testing has also used some form of Pauli
sampling\footnote{From hereon we drop the absolute values around $\alpha_P$, as
Hamiltonians are Hermitian and therefore have real Pauli coefficients.}
(normalized by $\norm{H}_2^2$ to give a probability distribution). One
approach, used in~\cite{AAEG25}, is to use short time evolution so that
$e^{-iHt} \approx I - iHt$. Then preparing $\ket{e^{-iHt}}$ and measuring will
give a sample from the Pauli spectrum of $H$ with probability $\norm{H}_2^2
t^2$.

Under the standard assumption that $\norm{H}_\infty \le 1$, this requires setting $t
= \bT{\eps}$ to obtain a sample within $\eps^2$ total variation distance of an
exact Pauli sample, as we would want for $\eps$-accuracy property testing. This
then requires $\bT{1/\eps^2\norm{H}_2^2}$ iterations and so
$\bT{1/\eps\norm{H}_2^2}$ total evolution time.

So if the complexity of testing a property via Pauli sampling goes as $f(\eps)$
in the unitary setting, this transformation results in a Hamiltonian algorithm
with evolution time complexity $tf(\eps t)$, as each iteration of sampling on
$e^{-iHt}$ takes $t$ time. In the general case, $t$ needs to be at most
$\bT{\eps}$ for the linear approximation to $e^{-iHt}$ to be good enough,
resulting in, e.g.\ an $\bO{1/\eps^2}$ time unitary property testing algorithm
becoming an $\bO{1/\eps^3}$ time Hamiltonian property testing algorithm.

The algorithm of~\cite{KL25} effectively achieves this in the special case
where Pauli sampling is only needed to identify whether the Hamiltonian has
support on some known set of ``bad'' Paulis $S$, by in effect filtering out the
elements of the Pauli spectrum of $H$ that are \emph{not} in $S$. However,
their method depends on knowing $S$, and is less effective in the ``tolerant''
testing case.

We give a method for Pauli sampling from an unknown Hamiltonian with only
polylogarithmic overhead in the desired total variation distance, allowing
unitary property testing methods to be generically applied to Hamiltonian
property testing.

\subsection{Our Results}
Our core result, used in all of our algorithms, is an evolution-time-efficient
means of block encoding $H$, i.e.\ constructing a unitary \[
  U = \begin{pmatrix}
    \wt{H}/\alpha & *\\
    * & *
  \end{pmatrix}
\]
for a known normalization factor $\alpha$, where $\wt{H}$ is $\eps$-close to
$H$ in normalized Frobenius norm. 
\begin{theorem}[Informal version of \cref{lem:block-encoding-H}]
  \label{thm:infblock}
  We can obtain an $\eps$-accurate block-encoding of an unknown Hamiltonian $H$
  with normalization factor $\bO{\log^2(1/\eps)}$ with $\bO{\eps^{-1}\log^5 (1/\eps)}$
  queries to its time evolution operator and $\bO{\log (1/\eps)}$ total
  evolution time.
\end{theorem}
As with all of our results, this requires neither controlled application of the
time evolution operator $e^{-iHt}$ nor reverse time evolution. 

As a corollary, we are able to sample from the Pauli spectrum of $H$ with
minimal overhead.
\begin{theorem}[Informal version of \cref{thm:spectrum-sampling-H}]
Write $H$ in the $n$-fold Pauli basis $\Pbb_n$ as \[
  \sum_{P \in \Pbb_n} \alpha_P P. 
\]
We can sample from the probability distribution on $\Pbb_n$ given by
$\alpha_P^2/\norm{H}_2^2$ with $\eps$ accuracy using $\bO{\frac{\log^9
(1/\eps)}{\eps \norm{H}_2^2}}$ queries to the time evolution operator and
$\bO{\frac{\log^5 (1/\eps)}{\norm{H}_2^2}}$ total evolution time per sample.
\end{theorem}
We use this sampling primitive to develop tolerant property testing algorithms
for locality, sparsity, degree-boundedness, being a $k$-junta, and Pauli
dimensionality, all scaling with $\frac{1}{(\eps_2 - \eps_1)^2}$.
\begin{theorem}[Informal version of
  \cref{thm:sparsity-tester-tolerant,thm:degree-boundedness-tester-tolerant,thm:juntaness-tolerant,thm:paulidimensionality-tolerant,thm:locality-tester-tolerant}]
  There are $(\eps_1,\eps_2)$-tolerant Hamiltonian property testers for:
  \begin{itemize}
    \item $k$-locality, using $\bOt{\frac{1}{\eps_2(\eps_2 - \eps_1)^3}}$
      queries to its time evolution operator and $\bOt{\frac{1}{(\eps_2 -
      \eps_1)^2}}$ total evolution time.
    \item $s$-sparsity, using $\bOt{\frac{s}{\eps_2(\eps_2 - \eps_1)^3}}$
      queries to its time evolution operator and $\bOt{\frac{s}{(\eps_2 -
      \eps_1)^2}}$ total evolution time.
    \item $\Delta$-degree-boundedness, using $\bOt{\frac{n\Delta}{\eps_2(\eps_2
      - \eps_1)^3}}$ queries to its time evolution operator and
      $\bOt{\frac{n\Delta}{(\eps_2 - \eps_1)^2}}$ total evolution time.
    \item Being a $k$-junta, using $\bOt{\frac{k}{\eps_2(\eps_2 - \eps_1)^3}}$
      queries to its time evolution operator and $\bOt{\frac{k}{(\eps_2 -
      \eps_1)^2}}$ total evolution time.
    \item Having Pauli dimensionality $k$, using $\bOt{\frac{k}{\eps_2(\eps_2 -
      \eps_1)^3}}$ queries to its time evolution operator and
      $\bOt{\frac{k}{(\eps_2 - \eps_1)^2}}$ total evolution time.
  \end{itemize}
\end{theorem}
For locality testing,~\cite{KL25} gave a tolerant tester using $\bOt{\frac{\eps_2^{1/2}}{(\eps_2 -
\eps_1)^{5/2}}}$ evolution time. For sparsity testing, the best non-tolerant
tester was given by~\cite{ST25}, achieving $\bOt{\frac{s^{3/2}}{\eps} +
\frac{s}{\eps^2}}$, while the best tolerant tester was the
$\bOt{\frac{s^5}{(\eps_2^2 - \eps_1^2)^5}}$ evolution time algorithm
of~\cite{AAEG25}. We are not aware of previous work explicitly considering the
other problems mentioned, but both~\cite{BCO24} and~\cite{AAEG25} give generic
testers for properties described in terms of a family of allowable Pauli
supports. The former of these would give
$\bOt{\frac{n\Delta}{\eps_2^{1/2}(\eps_2 - \eps_1)^{5/2}}}$ and
$\bOt{\frac{k}{\eps_2^{1/2}(\eps_2 - \eps_1)^{5/2}}}$ evolution time testers
for degree-boundedness and $k$-juntas, respectively.

Beyond its direct use as a sampler, we also use \cref{thm:infblock} to develop
a tolerant Hamiltonian \emph{certification} algorithm, in which the property
tested is identity to a given Hamiltonian $H_0$.
\begin{theorem}[Informal version of~\cref{thm:hamiltonian-certification}]
  There is an $(\eps_1,\eps_2)$-tolerant algorithm for certifying whether $H$
  is equal to a known Hamiltonian $H_0$, using $\bOt{\frac{1}{(\eps_2 -
  \eps_1)^3}}$ queries to its time evolution operator and
  $\bOt{\frac{1}{(\eps_2 - \eps_1)^2}}$ total evolution time.
\end{theorem}
Hamiltonian certification was originally considered by~\cite{GJWYZ26} in the
setting in which inverses (and therefore quantum amplitude estimation) are
allowed, in which they obtained $\bO{\frac{1}{\eps_2 - \eps_1}}$ evolution
time. Without inverses, the best known result from previous work is
$\bOt{\frac{\eps_2^{1/2}}{(\eps_2 - \eps_1)^{5/2}}}$, from applying the
tolerant emptiness tester of \cite{ST25} to $e^{-i(H - H_0)t}$ (in terms of
evolution \emph{time}, certification and emptiness testing are equivalent, as
$e^{-i(H - H_0)t}$ can be approximated to arbitrary accuracy with alternating
short-time evolutions under $e^{-iHt}$ and $e^{iH_0t}$).

The lower bounds of~\cite{GLZ26, allen2026power} establish
$\bOm{\frac{1}{\eps^2}}$ lower bounds for many of these problems in the
non-tolerant setting. We extend this (not quite tightly) to the tolerant
setting by a reduction to the distinguishing problem studied in~\cite{GLZ26}.
\begin{theorem}[Informal version of~\cref{thm:tolerant-property-testing-lower-bounds}]
  Any $(\eps_1,\eps_2)$-tolerant property tester for locality, sparsity,
  degree-boundedness, being a junta, Pauli dimensionality, or for certifying
  equality to a known $H_0$, requires $\bOm{\min\set*{\frac{1}{\eps_2(\eps_2 -
  \eps_1)}, \frac{2^{n/2}}{\sqrt{\eps_2(\eps_2 - \eps_1)}}}}$ total evolution
  time.
\end{theorem}

Our results all assume that $\norm{H}_\infty \le 1$. This is a standard assumption; the
one exception we are aware of is the $\bOt{\frac{s^{3/2}}{\eps} +
\frac{s}{\eps^2}}$ time sparsity testing algorithm of~\cite{ST25}, which
instead scales linearly with $\norm{H}_\infty$. We note that, as our algorithms
have inverse-quadratic scaling, this can be replicated for any of our results,
by running them on $H/\norm{H}_\infty$. We also follow the standard convention
of assuming $H$ is traceless, as in the absence of controlled evolution the
identity coefficient of $H$ can only cause an undetectable global phase shift.

Our algorithms use $3n + \plog(1/\eps)$ ancilla qubits. Of these, $n$ come from our
use of the Choi state, while the remainder are a consequence of our block-encoding
technique.

\subsection{Organization}
In \cref{sec:proof-overview}, we give a high-level overview of our proofs.
\cref{sec:block-encoding-spectrum-sampling} gives the full proof of our main
technical result: the construction of a block encoding of $H$ and the use of it
to sample from the Pauli spectrum of $H$.

\cref{sec:property-testing} applies this technique to derive tolerant
property testers for a number of properties of Hamiltonians.
\cref{sec:lower-bounds} gives the corresponding lower bounds. 

\section{Technical Overview}
\label{sec:proof-overview}
\subsection{Pauli Sampling via Block Encoding}
Our main technical tool is a method for block-encoding an unknown Hamiltonian
$H$ using its time evolution operator. That is, we construct a unitary \[
\begin{pmatrix}
  \wt{H}/\alpha & * \\
* & *
\end{pmatrix}
\] using applications of the time evolution operator, where $\wt{H}$
approximates $H$ and $\alpha$ is a known normalization factor. The most similar
prior work is that of~\cite{Z25}, which constructs the ``pseudo-Choi state''
$\ket{H}/\norm{H}_2$ by applying the linear combination of unitaries technique
to queries to the Hamiltonian time evolution operator. We improve the
approximation error substantially by using Chebyshev interpolation, also used
in~\cite{CY26} for a similar application, to substantially reduce the
coefficient size needed in the LCU application.

We start from the observation that \[
  H = i \left.\frac{d}{dt} e^{-iHt}\right|_{t=0}
\]
and so it would suffice to construct the \emph{derivative} of the time
evolution operator, evaluated at zero. This is not immediately useful, as we do
not have direct access to the derivative of the operator.  However, we know
that $e^{-iHt}$ can be $\eps$-approximated by the first $\bO{\log(1/ \eps)}$ terms
of its Taylor expansion, which is a degree-$\bO{\log(1/\eps)}$ polynomial, and
that (by standard tricks using Lagrange polynomials) \emph{any} degree-$k$
polynomial $p$ has \[
  p'(0) = \sum_{j=0}^k w_j p(t_j)
\]
for $t_j$ in $\brac{0,1}$ and $\abs{w_j}$ summing to $2k^2$. So if we could block-encode
the sum \[
  i\sum_{j=0}^k w_j e^{-iHt_j} 
\]
we could block-encode $H$. The natural strategy would be to use
the ``linear combination of unitaries'' technique from Hamiltonian simulation
to block-encode this sum. However, this technique requires
controlled access to the unitaries, which our access model does not allow.
Fortunately, a result of~\cite{Z25} shows that it is possible to
approximate controlled access to unitary time evolution using uncontrolled queries.

However, this technique constructs a channel $\gamma$-far from the controlled
access unitary in diamond norm distance at query cost proportional to
$1/\gamma$ (although without additional evolution time overhead), which would
inflate our query complexity by $1/\norm{H}_2^2$, as we would need to set
$\gamma$ proportional to $\norm{H}_2^2$ to avoid destroying the approximation
of $H$. So instead we apply the technique of~\cite{Z25} to construct a block
encoding of the controlled access unitary, which we then use in LCU to
construct a block-encoding of $H$.

The cost of this procedure is proportional to the total magnitude of the
coefficients in the LCU sum, which as we have seen is $\bO{\log^2 (1/\eps)}$.
This immediately allows Pauli sampling from the spectrum of $H$ (using $n$
additional ancilla qubits). Prepare \[
\begin{pmatrix}
  I \otimes \wt{H}/\alpha & * \\
* & *
\end{pmatrix} 
2^{-n/2}\sum_i \ket{i}\ket{i}\ket{0}
\]
and measure with a basis containing \[
\begin{pmatrix}
  \ket{P}\\
0
\end{pmatrix}
\]
for each Pauli $P$. With probability $\bT{\norm{\wt{H}/\alpha}_2^2}$ we will obtain a
sample from the Pauli basis of $\wt{H}$, and so $\bTt{1/\norm{H}}_2^2$
repetitions of the process per sample suffice.

\subsection{Application to Hamiltonian Property Testing}
Our testing algorithms follow substantively similar strategies to previous work
in Hamiltonian and unitary property testing. The ability to sample from the
Pauli basis of $H$ directly allows us to simplify and strengthen certain prior
Hamiltonian testing algorithms, and in other cases we directly adopt strategies
used for testing properties of unitaries.
\subsubsection{Locality}
Locality (or any other property defined as being supported on a prescribed
subset of Paulis) is particularly simple to test once given sampling access to
the Pauli spectrum of $H$: the distance of $H$ from the closest $k$-local
Hamiltonian is exactly the two-norm of its non-local Pauli coefficients.
Therefore, if $H$ is $\eps$-far from $k$-locality, the probability of sampling
a non-local Pauli is $\eps^2/\norm{H}_2^2$. Standard variance arguments
therefore tell us that $\bO{\frac{\norm{H}_2^2}{(\eps_2 - \eps_1)^2}}$ samples
suffice to test whether the distance of \(H\) from \(k\)-locality is \(\le
\eps_1\) or \(\ge \eps_2\), with total evolution time cost of
$\bOt{\frac{1}{(\eps_2 - \eps_1)^2}}$.

This is the same strategy used in~\cite{AAEG25,KL25}, and analogous to a
strategy used in \cite{BCO24}. Relative to~\cite{AAEG25} we improve from cubic
dependence on $(\eps_2 - \eps_1)^{-1}$ in our evolution time bound to
quadratic, because of the lower overhead of our sampling primitive. The
algorithm of~\cite{KL25} already achieved quadratic dependence via a
low-overhead sampling primitive based on ``Trotterized postselection''.
However, their method only preserves the Pauli mass of the non-local terms of
$H$ up to a constant multiplicative factor, which is sufficient for
\emph{non-tolerant} testing, but not tolerant testing. Preserving this
non-local mass to sufficient accuracy for tolerant testing increases the
overhead of their sampler, resulting in $\bO{\sqrt{\frac{\eps_2}{(\eps_2 -
\eps_1)^5}}}$ complexity.

\subsubsection{Sparsity and Degree-Boundedness}
We treat the properties of $s$-sparsity and $\Delta$-degree-boundedness as
instances of what we call ``generalized cardinality testing'', in which some
subset $\Sc$ of the powerset of the Paulis is associated with a size function
$\ell$, and we deem $H$ to have cardinality $k$ if its support is contained in
an $S \in \Sc$ with $\ell(S) \le k$. For the remainder of this overview, we
will discuss the sparsity case alone, as the generalization is not
substantially different.

Our algorithm follows a similar procedure to the tolerant sparsity tester
of~\cite{AAEG25}. We use Pauli sampling to find candidate elements for an
$s$-sparse support for $H$, and then sample enough Paulis to estimate how much
of the Pauli mass of $H$ is accounted for by the ``best'' support.

We improve on the algorithm of~\cite{AAEG25} in two ways. The first is our
low-overhead sampling algorithm, which accounts for an improvement of
$\bT{\frac{s}{\eps_2^2 - \eps_1^2}}$ relative to their
$\bO{\frac{s^5}{(\eps_2^2 - \eps_1^2)^5}}$ sparsity tester. The second is an
improved analysis of the number of samples needed to identify the best
candidate support and estimate the residual Pauli mass outside it.

Previous algorithms for this problem, including~\cite{AAEG25} and the
non-tolerant algorithm of~\cite{ST25}, implicitly or explicitly relied on
$(\eps_2^2 - \eps_1^2)$-accurately estimating the Pauli mass of the $s$
heaviest terms of $H$. However, this becomes a problem when it comes to the
\emph{heaviest} terms of $H$: in general, the complexity of estimating the
parameter $p$ of a Bernoulli random variable to accuracy $\delta$ goes as
$p/\delta^2 + 1/\delta$. We avoid this issue by taking advantage of the fact
that the heaviest Paulis in $H$ are required to be in any $s$-sparse support,
and therefore no precise estimate of their weight is needed to estimate the
distance of $H$ from sparsity.

We first take $\bOt{\frac{s}{\eps_2^2 - \eps_1^2}}$ Pauli samples $C$. This is
enough to guarantee, by a union bound, that every size-$s$ set of Paulis has at
most $\frac{\eps_2^2 - \eps_1^2}{10}$ mass outside of $C$. We can therefore
restrict our set of candidate $s$-sparse representations of $H$ to subsets of
$C$. Then, we use another round of sampling to estimate the Pauli mass outside
of every $S \subseteq C$ of size $s$, either returning an $\frac{\eps_2^2 -
\eps_1^2}{10}$-accurate estimate of the Pauli mass or certifying that it is at
least, say, $5\eps_2^2$. Any set where the latter holds cannot be a good
$s$-sparse support for $H$, so it then suffices to check every set where the
former holds to see if any gives an $s$-sparse support that is within $\eps_1$
of $H$.

The number of samples needed for the final procedure is, by a union bound over
the $\binom{\abs{C}}{s} = \bOt{\poly(\eps_2 - \eps_1)^{-s}}$ candidates and standard
results on Bernoulli estimation, \[
  \bO{s \cdot \plog\left(\frac{1}{\eps_2 - \eps_1}\right) \cdot \max\paren*{\frac{\eps_2^2}{(\eps_2^2 -
  \eps_1^2)^2}, \frac{1}{\eps_2^2 - \eps_1^2}}} = \bOt{\frac{s}{(\eps_2 -
  \eps_1)^2}}.
\]

\subsubsection{Juntas and Pauli Dimensionality}
Like with sparsity and degree-boundedness, we treat junta testing and Pauli
dimensionality testing as instances of what we call ``generalized
dimensionality testing'', in which we have a set of
\emph{subgroups}\footnote{As any set of subgroups is also a set of subsets, it
might initially seem that this should reduce to the generalized cardinality
testing problem in the previous section. However, our algorithms there use an
assumption that the ``size'' of a set is at least a constant factor times its
actual cardinality, whereas here we allow the dimension to be logarithmic in
the size of the subgroup.} of the Paulis, and a function $\ell$ describing the
``dimension'' of these subgroups, and the property to be tested is whether $H$
is contained in a subgroup of dimension at most $k$. 

Like in the previous section, we will describe the algorithm in the case of
testing $k$-juntas, as the generalization is very similar. Our algorithm is
very similar to prior work on testing unitaries~\cite{BLLYZ25} and Boolean
functions~\cite{C26}, although we are able to obtain stronger bounds than were
known in the unitary case (in part because the distance of a Hamiltonian from
being a $k$-junta is more directly characterized by its Fourier spectrum than
for unitaries). It is also morally quite similar to the tolerant sparsity
testing algorithm of the previous section.

Let $W$ be the $k$ qubits that minimize the weight of Paulis in $H$ acting
outside of $W$. We sample a size-$\bTt{\frac{k}{\eps_2^2 - \eps_1^2}}$ set of
Paulis $C$, which suffices to guarantee that there is some subset $C'$ of our
$C$ such that $\abs{C'} \le k$, $C'$ acts on at most $k$ qubits, and there is
at most $\frac{\eps_2^2 - \eps_1^2}{10}$ total weight of terms in $H$ that only
act on qubits in $W$ but act on some qubits not acted upon by $C'$.

This means we only have $\binom{\abs{C}}{k}$ candidate sets of qubits to check,
even though in general there could be as many as $\binom{n}{k}$, (even if we
restricted to qubits acted upon by at least one sampled term). So as in the
sparsity testing case, it will suffice to take \[
  \bOt{\frac{k}{(\eps_2 - \eps_1)^2}} 
\]
samples to test every candidate set of $k$ qubits, either estimating the mass
on terms acting outside it to $\frac{\eps_2^2 - \eps_1^2}{10}$ accuracy or
certifying that this mass is a large enough multiple of $\eps_2^2$.

\subsubsection{Hamiltonian Certification}
Our tolerant Hamiltonian certification algorithm is the only one of our
algorithms that does not directly proceed from our sampling result. Instead, we
use our block encoding result, together with another application of the linear
combination of unitaries technique, to block-encode $H - H_0$ and then either
certify that $\norm{H - H_0}_2 > \eps_2$ or estimate $\norm{H - H_0}_2^2$
to additive $\eps_2^2 - \eps_1^2$ error by sampling. This requires \[
\bO{\max\paren*{\frac{\eps_2^2}{(\eps_2^2 - \eps_1^2)^2}, \frac{1}{\eps_2^2 -
\eps_1^2}}} = \bO{\frac{1}{(\eps_2 - \eps_1)^2}}
\]
samples. By comparison, the algorithm of~\cite{GJWYZ26} uses quantum amplitude
estimation on a simulation of evolution under $H - H_0$, but this requires access to inverse time
evolution.

Another very similar approach that is implicit in prior work is applying the
tolerant emptiness tester of~\cite{ST25} to $e^{-i(H - H_0)t}$ (which can be
approximated by alternating short-time applications of $e^{-iHt}$ and
$e^{iH_0t}$). This would give
complexity of \[ 
  \bO{\frac{\eps_2^{1/2}}{(\eps_2 - \eps_1)^{5/2}}}
\]
total evolution time. The gap here exists for the same reason as in the
locality testing case: \cite{ST25} can sample without overhead if they are
willing to disturb the distribution of their samples by a constant factor
(which is acceptable when $\eps_2 - \eps_1 = \bOm{\eps_2}$) but the precision
needed to estimate $\norm{H - H_0}_2$ to $\eps_2 - \eps_1$ accuracy increases
their overhead when $\eps_2 - \eps_1$ is smaller.

\subsection{Lower Bounds}
Our lower bounds are a direct extension of the non-tolerant locality testing
lower bounds of~\cite{GLZ26}. They gave a distribution on Hamiltonians that is
$\bOm{\eps}$ far from being $k$-local, while requiring
$\bOm{\min(1/\eps^2,2^{n/2}/\eps)}$ total evolution time to distinguish from
the zero Hamiltonian. To give lower bounds for our problem we need to show that
this Hamiltonian is far from being sparse, low-degree, a junta, or of low Pauli
dimensionality. As in the case of locality, this is a direct consequence of the
Hamiltonian distribution having support spread across the entire diagonal.

This gives $\bOm{1/\eps^2}$ lower bounds for non-tolerant testing. To extend
this to the $(\eps_1,\eps_2)$-tolerant setting, we add an additional
Hamiltonian that is fixed and $\eps_1$-far from having the property. Our lower
bounds then follow by setting $\eps = \bT{\sqrt{\eps_2^2 - \eps_1^2}}$.

\section{Preliminaries}
\label{sec:preliminaries}

A Hamiltonian \(H\) on \(n\) qubits is a \(2^n \times 2^n\) Hermitian matrix. The time evolution operator of a Hamiltonian \(H\) for time \(t \geq 0\) is the unitary matrix
\begin{align*}
    e^{-iHt} \coloneq \sum_{j = 0}^{\infty} \frac{(-iHt)^j}{j!}.
\end{align*}
Our algorithms assume query access to this time evolution operator for any time \(t\).  We write \(\ket\psi\) for a quantum state vector and \(\bra\psi\) for its conjugate transpose.  If a state is expanded in an orthonormal basis as \(\sum_y a_y\ket y\), then measuring in that basis returns outcome \(y\) with probability \(\abs{a_y}^2\).  This is called Born's rule and is the only measurement rule used in our algorithms.

\subsection{Distance measures}

We define the \(n\)-qubit Pauli matrices to be \(\Pbb_n = \{I, X, Y, Z\}^{\otimes n}\) where
\begin{align*}
    X = \begin{pmatrix}
        0 & 1 \\ 1 & 0
    \end{pmatrix},
    Y = \begin{pmatrix}
        0 & -i \\ i & 0
    \end{pmatrix},
    Z = \begin{pmatrix}
        1 & 0 \\ 0 & -1
    \end{pmatrix}.
\end{align*}
The Pauli strings form an orthogonal basis for the space of \(2^n\times2^n\) matrices.  Thus every operator \(M\) has a unique expansion
\begin{align*}
    M=\sum_{P\in\mathcal{P}_n}\widehat M(P)P,
    \quad \text{where}~\widehat M(P)=2^{-n}\Tr(PM).
\end{align*}
We write $\supp(H)$ for the set of Paulis $P$ for which $\wh{H}(P) \neq 0$.

For an \(n\)-qubit operator \(A = \sum_{P \in \Pbb_n} \alpha_P P\), the normalized Frobenius norm is given by
\begin{align*}
    \norm{A}_2 = \sqrt{\frac{\Tr{(A^\dagger A)}}{2^n}} = \sqrt{\sum_{P \in \Pbb_n} \abs{\alpha_P}^2}.
\end{align*}
The normalization by \(2^{-\frac{n}{2}}\) makes this norm dimension-independent
on bounded Hamiltonians. The spectral norm (Schatten-\(\infty\) norm),
\(\norm{A}_\infty\), is the largest singular value of \(A\). Throughout the
work, for any Hamiltonian \(H\) we assume \(\norm{H}_\infty \le 1\), as a form
of normalization. We also assume that \(\Tr(H)=0\). This assumption is without
loss of generality: the trace of a Hamiltonian changes its time evolution only
by a global phase, which cannot affect our algorithms, as they do not have
access to controlled time evolution. Our properties are defined to not depend
on the trace of $H$. While removing the identity coefficient of $H$ can
increase its operator norm, it can only do so by a constant factor, which would
not affect any of our results.

The diamond norm of a linear map \(\Mc\) is
\begin{align*}
    \norm{\Mc}_\diamond
    =\sup_{\substack{m\ge1\\X\neq0}}
    \frac{\norm{(\Mc\otimes\operatorname{id}_m)(X)}_1}
    {\norm{X}_1},
\end{align*}
where \(\operatorname{id}_m\) is the identity channel on an \(m\)-dimensional ancillary system, and \(\norm{X}_1=\Tr\left(\sqrt{X^\dagger X}\right)\).

We will repeatedly use the following facts in our proofs.
\begin{fact}[Non-commutative telescoping identity]\label{fact:telescoping-identity}
    For any two matrices \(A\) and \(B\)
    \[
      A^j - B^j = \sum_{r = 0}^{j - 1} A^r (A - B) B^{j - 1 - r}.
    \]
\end{fact}

\begin{fact}\label{fact:infinity-norm-power}
    For any operator \(A\) such that \(\norm{A}_\infty \le 1\), for all \(j \in
    \Zbb_{\ge 2}\), \(\norm{A^j}_\infty \le \norm{A}_\infty^j\).
\end{fact}

\begin{fact}[Taylor's formula]\label{fact:taylor}
    For any Hamiltonian \(H\), \(k\in\Zbb_{\ge0}\), and \(t\ge0\),
    \begin{align*}
        e^{-itH}
        =\sum_{j=0}^k\frac{(-itH)^j}{j!}
        +\frac{(-iH)^{k+1}}{k!}
        \int_0^t(t-u)^ke^{-iuH}\,du.
    \end{align*}
\end{fact}

\subsection{Choi states}

Let
\begin{align*}
    \ket{\Phi}
    \coloneqq
    2^{-\frac{n}{2}}\sum_{x\in\{0,1\}^n}\ket{x}\ket{x}
\end{align*}
be the maximally entangled state on two \(n\)-qubit registers. 

\begin{definition}
    For any unitary \(U \in \Cbb^{2^n \times 2^n}\), define
    \begin{align*}
        \ket{U}
        \coloneqq
        (I^{\otimes n}\otimes U)\ket{\Phi}.
    \end{align*}
    These states are (a special case of) \emph{Choi states}. The isomorphism between the Pauli strings and the Choi states is known as the \emph{Choi-Jamio\l kowski isomorphism}~\cite{Jam72, Choi75}. The isomorphism is simply \(U \leftrightarrow \ket{U} \).
\end{definition}

The states \(\{\ket{P}:P\in\mathcal{P}_n\}\) form an orthonormal basis.  We refer to this as the \emph{Bell basis}.

\begin{lemma}\label{lem:bell-sampling}
    Given a unitary \(U=\sum_{P\in\mathcal{P}_n}\widehat U(P)P\), we can sample a Pauli $P$ with probability \(\abs{\widehat U(P)}^2\).
\end{lemma}

\begin{proof}
    Prepare \(\ket{\Phi}\), apply \(I^{\otimes n}\otimes U\), and measure in the Bell basis.
\end{proof}

\begin{definition}
    For a non-zero, \(n\)-qubit operator \(A\), the \emph{(referenceless) pseudo-Choi state} of \(A\), as defined in~\cite{Z25}, is the state
    \begin{align*}
        \ket{\Psi(A)} \coloneq \frac{(I \otimes A) \ket{\Phi}}{\norm{(I \otimes A) \ket{\Phi}}}.
    \end{align*}
\end{definition}
Also observe that \(\norm{(I \otimes A) \ket{\Phi}} = \norm{A}_2\).

\subsection{Block encodings}

\begin{definition}\cite[Definition 43]{GSLW19}
    Let \(A\) be an \(n\)-qubit operator and \(\alpha, \eps \ge 0\). An \((n + a)\)-qubit unitary operator \(B\) is called an \((\alpha, a, \eps)\)-\emph{block encoding} of A if
    \begin{align*}
        \norm{A - \alpha(\bra{0^a} B \ket{0^a})}_\infty \le \eps.
    \end{align*}
\end{definition}

For brevity, we will use the shorthand: \((\alpha, a, \eps)\)-BE. For any unitary \(B\) which is an \((\alpha, a, \eps)\)-BE of some operator \(A\), we refer to \((\bra{0^a}\otimes I)B(\ket{0^a}\otimes I)\) as the \emph{zero block} of \(B\).

\begin{definition}\cite[Definition 51]{GSLW19}
    Let \(y \in \Cbb^K\) with \(\norm{y}_1 \le \beta\) and \(\eps \ge 0\). The pair of \(b\)-qubit unitaries \((\texttt{PREP}_L, \texttt{PREP}_R)\) is called a \((\beta, b, \eps)\)-\emph{state preparation pair} for \(y\) if
    \begin{align*}
        \sum_{j = 0}^{K - 1} \abs{\beta(c_j^*d_j) - y_j} \le \eps,
    \end{align*}
    where \(c_j = \bra{j} \texttt{PREP}_L \ket{0^b}\) and \(d_j = \bra{j} \texttt{PREP}_R \ket{0^b}\), and for all \(j = K, \dots, 2^b - 1\), \(c_j^*d_j = 0\).
\end{definition}

For brevity, we will use the shorthand: \((\beta, b, \eps)\)-SP pair. 

\begin{lemma}\cite[Lemma 52]{GSLW19}\label{lem:linear-combination-be}
    Let \(A = \sum_{j = 0}^{K - 1} y_j A_j\) be an \(n\)-qubit operator, expressed as a linear combination of \(K\) operators \(A_j\) with \(y \in \Cbb^K\). Let \(\eps_1, \eps_2 \ge 0\). For each \(A_j\), suppose that there is an \((\alpha, a, \eps_2)\)-block encoding \(B_j\). Also, let \((\normalfont{\texttt{PREP}}_L, \normalfont{\texttt{PREP}}_R)\) be a \((\beta, b, \eps_1)\)-state preparation pair for \(y\). Define the \((n + a + b)\)-qubit unitary
    \begin{align*}
      \normalfont{\texttt{SEL}} = \sum_{j = 0}^{K - 1} \ket{j}\bra{j} \otimes B_j + \sum_{j = K}^{2^b - 1} \ket{j}\bra{j} \otimes I^{\otimes (n + a)}.
    \end{align*}
    Then, \((\normalfont{\texttt{PREP}}_L^\dagger \otimes I^{\otimes (n + a)}) \normalfont{\texttt{SEL}} (\normalfont{\texttt{PREP}_R} \otimes I^{\otimes (n + a)})\) is an \((\alpha \beta, a + b, \alpha \eps_1 + \beta \eps_2)\)-block encoding of \(A\).\footnote{A typo in~\cite{GSLW19} mistakenly carries a factor of \(\alpha\) in the \(\beta \eps_2\) error term.}
\end{lemma}

\subsection{Lagrange interpolation and Chebyshev polynomials}

\begin{fact}[Chebyshev polynomials of the first kind]\label{fact:chebyshev-polynomials}
    For any \(k\in\Zbb_{\ge0}\), the Chebyshev polynomial of the first kind, \(T_k\), is the unique polynomial of degree \(k\) satisfying
    \[
        T_k(\cos\theta)=\cos(k\theta),
        \qquad \theta\in[0,\pi].
    \]
\end{fact}

\begin{fact}[Derivative of Chebyshev polynomials]\label{fact:chebyshev-derivative}~\cite[Example 3.2.3]{dahlquist2008numerical}
    For any \(k\in\Zbb_{\ge1}\), the Chebyshev polynomial of the first kind, \(T_k\), satisfies
    \[
        T_k'(1)=k^2.
    \]
\end{fact}

\begin{fact}[Lagrange interpolation]\label{fact:lagrange-interpolation}
    For \(k\in\Zbb_{\ge1}\) and distinct real numbers \(t_0,\ldots,t_k\), let
    \[
        L_j(x)=\prod_{\substack{0\le r\le k\\r\ne j}}
        \frac{x-t_r}{t_j-t_r}.
    \]
    For any polynomial \(p\) of degree at most \(k\),
    \[
        p(x)=\sum_{j=0}^k p(t_j)L_j(x),
    \] and
    \[
        p'(0)=\sum_{j=0}^k p(t_j)L_j'(0).
    \]
\end{fact}

\begin{proof}
    The first equation is the well-known Lagrange interpolation formula (see~\cite[Equation 5.1]{trefethen2019approximation}). Differentiating and substituting \(x=0\) gives the second identity.
\end{proof}

\subsection{Probability}

We will use the following elementary concentration bounds.

\begin{lemma}[Hoeffding bound~\cite{hoeffding1963probability}]\label{lem:hoeffding-prelim}
    Let \(X_1,\ldots,X_N\in[0,1]\) be independent random variables with
    \(\E{X_i} = p\). Let
    \(
        \hat p=\frac{1}{N}\sum_{i=1}^N X_i.
    \) Then for every \(\eta>0\),
    \begin{align*}
        \prob{\abs{\hat p - p} > \eta}
        \le
        2\exp(-2N\eta^2).
    \end{align*}
    Consequently, if
    $
        N\ge \frac{1}{2\eta^2}\log\frac{2}{\delta},
    $ then \(\abs{\hat p-p}\le\eta\) with probability at least \(1-\delta\).
\end{lemma}

\begin{lemma}[Chernoff bound~\cite{chernoff1952measure}]\label{lem:chernoff-prelim}
    Let \(X_1,\ldots,X_N\in[0,1]\) be independent random variables with
    \(\E{X_i}=p\). Let
    $
        \hat p=\frac{1}{N}\sum_{i=1}^N X_i.
    $ Then for every \(\gamma\in(0,1]\),
    \begin{align*}
        \prob{\hat p\ge (1+\gamma)p}
        &\le
        \exp\left(-\frac{\gamma^2Np}{3}\right),\\
        \prob{\hat p\le (1-\gamma)p}
        &\le
        \exp\left(-\frac{\gamma^2Np}{2}\right).
    \end{align*}
    Consequently,
    \begin{align*}
        \prob{\abs{\hat p-p}>\gamma p}
        \le
        2\exp\left(-\frac{\gamma^2Np}{3}\right).
    \end{align*}
\end{lemma}

\begin{lemma}[Bernstein bound]
\label{lem:bernstein-prelim}
    Let \(X_1,\ldots,X_N\in[0,1]\) be independent random variables with
    \(\E{X_i}=p\). Let
    \(
        \hat p=\frac1N\sum_{i=1}^N X_i.
    \)
    Then for every \(\eta>0\),
    \begin{align*}
        \prob{\abs{\hat p-p}>\eta}
        \le
        2\exp\left(
            -\frac{N\eta^2}{2p+2\eta/3}
        \right).
    \end{align*}
    Consequently, if \(p\le v\) and
    \(
        N
        \ge
        \frac{2v+2\eta/3}{\eta^2}
        \log\frac{2}{\delta},
    \)
    then \(\abs{\hat p-p} \le \eta\) with probability at least \(1-\delta\).
\end{lemma}

\section{Block Encoding and Spectrum Sampling}
\label{sec:block-encoding-spectrum-sampling}

In this section, we give a method for block-encoding an unknown Hamiltonian $H$ via queries to its time evolution operator.
\begin{restatable}{theorem}{blockencodingH}\label{lem:block-encoding-H}
    For any Hamiltonian \(H\) such that \(\norm{H}_\infty \le 1\), and
    \(\gamma\in(0,1)\), we can construct a
    \(\left(\bO{\log^2\left(\frac{2}{\gamma}\right)}, \bO{n +
    \log{\left(\frac{2}{\gamma}\right)}}, \gamma\right)\)-block encoding of
    \(H\) with zero block $\wt{H}/\Lambda$ satisfying $\norm{\wt{H} - H}_2 \le
    \gamma \norm{H}_2$. This requires
    \(\bO{\frac{\log^5\left(\frac{2}{\gamma}\right)}{\gamma}}\) queries to the
    time evolution operator and \(\bO{\log\left(\frac{2}{\gamma}\right)}\)
    total evolution time.
\end{restatable}

As a consequence, we will show that it is possible to sample from a
distribution arbitrarily close to the Pauli spectrum of \(H\).

\begin{restatable}{theorem}{spectrumsampling}\label{thm:spectrum-sampling-H}
    For any Hamiltonian \(H\) such that \(\norm{H}_\infty \le 1\) and \(\gamma, \delta \in (0, 1/2)\), we can sample from a distribution \(\wh\Dc\) such that
    \[
        \norm{\wh\Dc-\Dc}_{\mathrm{TV}}\le\gamma,
        \qquad \text{where}~
        \Dc(P)=\frac{\alpha_P^2}{\norm{H}_2^2}.
    \]
    This requires, with probability at least \(1 - \delta\), \(\bO{\frac{\log^9\left(\frac{2}{\gamma}\right)}{\norm{H}_2^2\gamma}\log{\left(\frac1\delta\right)}}\) queries to the time evolution operator and \(\bO{\frac{\log^5\left(\frac{2}{\gamma}\right)}{\norm{H}_2^2}\log{\left(\frac1\delta\right)}}\) total evolution time.
\end{restatable}

We first present our block-encoding procedure. We then show how this lets us
prepare a (referenceless) pseudo-Choi state \(\ket{\Psi(\widetilde H)}\) with
\(\bOt{\frac1{\norm{H}_2^2}}\) total evolution time, where \(\norm{\widetilde H
- H}_\infty \le \gamma\) for any \(\gamma > 0\). Finally, we show that the
ability to prepare this state lets us sample from the Pauli spectrum of \(H\)
with arbitrary precision, using only \(\bOt{\frac{1}{\norm{H}_2^2}}\) total
evolution time. Throughout this section, we assume \(H \neq 0\).

We first show, in~\cref{cor:approximation}, that a Hamiltonian \(H\) can be approximated by a linear combination of \(\bOt{\log\left(\frac{2}{\gamma}\right)}\) time evolution operators, each having \(\bO{1}\) time evolution.

\begin{lemma}\label{lem:interpolation}
    For any Hamiltonian \(H\) such that \(\norm{H}_\infty\le1\), and \(k\in\Zbb_{\ge1}\), there exist real weights \(\set{w_j}_{j=0}^k\) and times \(0=t_0<\cdots<t_k=1\), such that \(\sum_{j=0}^k\abs{w_j}=2k^2\) and
    \begin{align*}
        \norm*{i\sum_{j=0}^kw_je^{-it_jH}-H}_\infty
        &\le\frac{2k^2}{(k+1)!},\\
        \norm*{i\sum_{j=0}^kw_je^{-it_jH}-H}_2
        &\le\frac{2k^2}{(k+1)!}\norm{H}_2.
    \end{align*}
\end{lemma}

\begin{proof}
    Let \(q(t)=\sum_{r=0}^k\frac{(-itH)^r}{r!}\), so that \(H=iq'(0)\). We will express \(q'(0)\) as a weighted sum of the values \(q(t_j)\), for appropriately chosen \(t_j\)'s, and then approximate each \(q(t_j)\) by \(e^{-it_jH}\).

    Take \(t_j=\frac{1-\cos(\frac{j\pi}{k})}{2}\) and \(w_j=L_j'(0)\), where \(L_j\) are the corresponding Lagrange basis polynomials. Applying Fact~\ref{fact:lagrange-interpolation} entrywise to \(q(t)\), we have that
    \[
        \sum_{j=0}^k w_jq(t_j)=q'(0)=-iH.
    \]
    Hence,
    \[
        i\sum_{j=0}^k w_je^{-it_jH}-H
        =i\sum_{j=0}^k w_j\left(e^{-it_jH}-q(t_j)\right).
    \]
    Therefore, it suffices to bound the Taylor remainder and the sum \(\sum_{j=0}^k\abs{w_j}\).

    Fact~\ref{fact:taylor} gives
    \[
        e^{-itH}-q(t)=\sum_{r=k+1}^\infty\frac{(-itH)^r}{r!}
        =\frac{(-iH)^{k+1}}{k!}
        \int_0^t(t-u)^ke^{-iuH}\,du.
    \]
    For \(t\in[0,1]\), the operator norm and the normalized Frobenius norm of this remainder are at most \(\frac1{(k+1)!}\) and \(\frac{\norm{H}_2}{(k+1)!}\), respectively, using \(\norm{H}_\infty\le1\) and
    \(\norm{H^{k+1}}_2\le\norm{H}_\infty^k\norm{H}_2\).

    To bound \(\sum_{j=0}^k\abs{w_j}\), let \(h(t)=T_k(1-2t)\), where \(T_k\) is the degree \(k\) Chebyshev polynomial of the first kind. By Facts~\ref{fact:chebyshev-polynomials}~and~\ref{fact:chebyshev-derivative}, we have \(h(t_j)=(-1)^j\) and \(h'(0)=-2k^2\). The definition of \(L_j\) gives \(w_0<0\) and \(\operatorname{sgn}(w_j)=(-1)^{j+1}\) for \(j\ge1\). Applying Fact~\ref{fact:lagrange-interpolation} to \(h\), we obtain
    \[
        -2k^2=\sum_{j=0}^k w_j(-1)^j
        =-\sum_{j=0}^k\abs{w_j}.
    \]
    Combining these estimates and using the triangle inequality gives
    \begin{align*}
        \norm*{i\sum_{j=0}^k w_je^{-it_jH}-H}_\infty
        &\le\frac{2k^2}{(k+1)!},\\
        \norm*{i\sum_{j=0}^k w_je^{-it_jH}-H}_2
        &\le\frac{2k^2\norm{H}_2}{(k+1)!}.
    \end{align*}
\end{proof}

\begin{corollary}\label{cor:approximation}
    For any Hamiltonian \(H\) such that \(\norm{H}_\infty\le1\), and \(\gamma \in (0,1)\), there exist \(k=\bO{\log\left(2 \gamma^{-1}\right)}\), real weights \(\set{w_j}_{j=0}^k\), and times \(0 = t_0<\cdots<t_k=1\), such that \(\sum_{j=0}^k \abs{w_j}= \bO{\log^2\left(2 \gamma^{-1}\right)}\) and
    \begin{align*}
        \norm*{i\sum_{j=0}^kw_je^{-it_jH}-H}_\infty
        \le\gamma,\\
        \norm*{i\sum_{j=0}^kw_je^{-it_jH}-H}_2
        \le\gamma \norm{H}_2.
    \end{align*}
\end{corollary}

\begin{proof}
    Let \(a_k = \frac{2k^2}{(k+1)!}\). For \(k < 3\), \(a_k \le \frac{4}{3}\). For \(k \ge 3\), we have \(
        \frac{a_{k + 1}}{a_k} = \frac{(k+1)^2}{k^2(k+2)}\le \frac12. \)
    
    Therefore,
    \(
        \frac{2k^2}{(k+1)!}\le6\cdot2^{-k}
    \) for all \(k\). Taking \(k=\max\set*{3,\ceil*{\log_2\left(\frac{6}{\gamma}\right)}}\), the result follows from \cref{lem:interpolation}.
\end{proof}

To construct our block encoding for \(H\), we will need controlled time evolution operations. We follow the notation of~\cite{Z25} for controlled unitary operations.

\begin{definition}\cite{Z25}
    For any bit string \(b \in \{0, 1\}^k\), the \(k\)-qubit controlled variant of the \(n\)-qubit unitary operator \(U\) is
    \begin{align*}
        \texttt{ctrl}_b(U) \coloneqq \ket{b}\bra{b} \otimes U + (I^{\otimes k} - \ket{b}\bra{b}) \otimes I^{\otimes n}
    \end{align*}
\end{definition}

We need the following lemma to construct our block encoding of the controlled time evolution operation.

\begin{lemma}\cite[Lemma 3]{Fang_2023}\label{lem:compression-gadget}
    Suppose we are given unitaries \(V_1, V_2, \ldots, V_L\), each of which is an \((\alpha'_l, m'_l, 0)\)-block encoding of \(\Gamma_l\). Then we can construct an \((\alpha_{\mathrm{comp}}, m_{\mathrm{comp}}, 0)\)-block encoding of \(\Gamma_L \cdots \Gamma_2 \Gamma_1\), where
    \[
        \alpha_{\mathrm{comp}} = \alpha'_1 \alpha'_2 \cdots \alpha'_L,
        \qquad
        m_{\mathrm{comp}} = \max_l m'_l + \left\lceil \log_2(L) \right\rceil + 1,
    \]
    using one application of each \(V_l\).
\end{lemma}

Using a coherent version of Pauli averaging, inspired by~\cite[Section 7]{Z25}, and using~\cref{lem:compression-gadget}, we obtain the following result on obtaining a block encoding of the controlled time evolution operation of a Hamiltonian.

\begin{lemma}\label{lem:controlization}
    For any Hamiltonian \(H\) such that \(\norm{H}_\infty\le1\), \(b \in \{0,1\}^k\), \(t \ge 0\),
    and \(\gamma\in(0,1)\), we can construct a
    \(\left(1, \bO{n + \log{\left(1 + \frac{t^2}{\gamma}\right)}}, \gamma\norm{H}_2^2 \right)\)-block encoding of \(\normalfont{\texttt{ctrl}}_b(e^{-itH})\). This requires \(\bO{1+\frac{t^2}{\gamma}}\) queries to the time evolution operator and \(t\) total evolution time.
\end{lemma}

\begin{proof}
    Consider a system with the following registers: a \(2n\)-qubit work register \(\Ac\), another \(n\)-qubit work register \(\Bc\), and a \(k\)-qubit control register \(\Cc\).
    
    Index the Paulis by \(P_0,\ldots,P_{4^n-1}\), and define the operations
    \[
        V_j=(\ket b\bra b)_{\Cc}\otimes (I)_{\Bc}
        +(I-\ket b\bra b)_{\Cc}\otimes (P_j)_{\Bc}
    \]
    for all \(j \in \{0, \dots, 4^n - 1\}\). \(V_j\) applies the Pauli \(P_j\) when the control register is not \(b\).

    For any duration \(v\ge0\), consider the unitary \(W(v)\) that does the following:
    \begin{itemize}
        \item Apply \(H^{\otimes 2n}\) to \(\Ac\) (preparing \(2^{-n}\sum_{j \in \{0,1\}^{2n}}\ket j\) from \(\ket{0^{2n}}\))
        \item Apply the operation \(\sum_{j \in \{0,1\}^{2n}}(\ket j\bra j)_{\Ac}\otimes V_j\), then the time evolution operation \(e^{-ivH}\) on \(\Bc\), and then the operation \(\sum_j(\ket j\bra j)_{\Ac}\otimes V_j\).
        \item Apply \(H^{\otimes 2n}\) to \(\Ac\).
    \end{itemize} 
    The zero block of the unitary \(W(v)\) is
    \begin{align*}
        (\bra{0^{2n}}\otimes I)W(v)(\ket{0^{2n}}\otimes I)
        &=\frac1{4^n}\sum_jV_j(I\otimes e^{-ivH})V_j\\
        &=\ket b\bra b\otimes e^{-ivH}
        +(I-\ket b\bra b)\otimes z(v)I,
    \end{align*}
    where \(z(v)=2^{-n}\Tr(e^{-ivH})\). The last equality follows from Pauli averaging: for any unitary \(U\) on \(n\) qubits
    \begin{align*}
        \frac{1}{4^n} \sum_{P \in \Pbb_n} PUP = \frac{\Tr(U)}{2^n} I^{\otimes n}
    \end{align*}
    Using Fact~\ref{fact:taylor}, \(\Tr(H)=0\), and
    \(\abs{\Tr(H^2e^{-iuH})}\le\Tr(H^2)\), we obtain
    \[
        \abs{z(v)-1}
        \le2^{-n}\int_0^v(v-u)\abs{\Tr(H^2e^{-iuH})}\,du
        \le\frac{v^2}{2}\norm{H}_2^2.
    \]
    Also, \(\abs{z(v)}\le1\).

    Set \(R=\max\set*{1,\ceil*{\frac{t^2}{2\gamma}}}\) and \(v=\frac{t}{R}\). By~\cref{lem:compression-gadget}, we can construct a \((1, 2n + \ceil*{\log(R)} + 1, 0)\)-block encoding of \[(\ket b\bra b\otimes e^{-ivH}
    +(I-\ket b\bra b)\otimes z(v)I)^R = \ket b\bra b\otimes e^{-itH} +(I-\ket b\bra b)\otimes (z(v))^RI.\]
    Call this block encoding \(B\). The distance of the zero block in \(B\) from \(\texttt{ctrl}_b(e^{-itH})\) is at most
    \[
        \abs{z(v)^R-1}
        \le R\abs{z(v)-1}
        \le\frac{t^2}{2R}\norm{H}_2^2
        \le\gamma\norm{H}_2^2.
    \]
    
    The circuit for \(B\) uses \(R = \bO{1 + \frac{t^2}{2\gamma}}\) queries to the time evolution operator with total evolution time \(Rv=t\).
\end{proof}

Before we describe how we block encode \(H\), we need the following lemma on multiplying the block encodings of multiple unitaries.

\begin{lemma}\cite[Corollary 55]{GSLW19}\label{lem:multiplying-unitary-be}
    Suppose that \(U_j\) is an \((1, a, \varepsilon)\)-block-encoding of an \(s\)-qubit unitary operator \(W_j\) for all \(j \in [K]\). Then \(\prod_{j=1}^K U_j\) is an \((1, a, 4K^2\varepsilon)\)-block-encoding of \(\prod_{j=1}^K W_j\).
\end{lemma}

This lemma is different from~\cref{lem:compression-gadget} in two ways. First, this lemma allows us to efficiently compose block encodings of unitaries, while~\cref{lem:compression-gadget} allows the composition of block encodings of operators that need not be unitary. Second,~\cref{lem:compression-gadget} composes block encodings that have no error, which is not the case in~\cref{lem:multiplying-unitary-be}.

We are now ready to describe our block encoding procedure. The procedure combines ~\cref{lem:linear-combination-be},~\cref{cor:approximation}, and~\cref{lem:controlization} to obtain the block encoding. We show that our block encoding of the \(\texttt{SEL}\) operator has error small enough to achieve the linear combination of unitaries with the desired accuracy, using which we achieve the following result.

\blockencodingH*

\begin{proof}
    Choose weights and times using~\cref{cor:approximation} such that the infinity-norm accuracy is \(\frac{\gamma}{2}\), and write
    \[
        B=i\sum_{j=0}^kw_je^{-it_jH},~\text{and}~
        \Lambda=\sum_j\abs{w_j}=2k^2.
    \]
    Then, by~\cref{cor:approximation}, \(k=\bO{\log\left(\frac{2}{\gamma}\right)}\), \(\norm{B-H}_\infty\le\frac{\gamma}{2}\), and
    \(\norm{B-H}_2\le\left(\frac{\gamma}{2}\right)\norm{H}_2\).

    For each \(1\le j\le k\), using~\cref{lem:controlization} obtain a \((1, a, \frac{\gamma}{8\Lambda k^2})\)-block encoding of \(\texttt{ctrl}_j(e^{-it_jH})\) (here \(j\) denotes its binary encoding), for some \(a\) depending on the accuracy parameter \(\frac{\gamma}{8\Lambda k^2}\). Denote the resulting block-encoding unitary by \(U_j\), and its zero block by \(C_j\). Therefore, by~\cref{lem:controlization}
    \[
        \norm{C_j-\texttt{ctrl}_j(e^{-it_jH})}_\infty
        \le\frac{\gamma}{8\Lambda k^2}\norm{H}_2^2,\\
        \qquad \norm{C_j}_\infty\le1.
    \]
    The exact controlled operations, \(\texttt{ctrl}_j(e^{-it_jH})\), composed together give us
    \[
        \texttt{SEL}=\sum_{j=0}^k
        \ket j\bra j\otimes e^{-it_jH} + \left(I - \sum_{j=0}^k
        \ket j\bra j\right)\otimes I.
    \] 
    By~\cref{lem:multiplying-unitary-be}, the unitary \(U_k\cdots U_1\) is a \(\left(1,a,\frac{\gamma}{2\Lambda}\norm{H}_2^2\right)\)-block encoding of \(\texttt{SEL}\). Denoting its zero block by \(\widetilde{\texttt{SEL}}\), we have
    \[
        \norm{\widetilde{\texttt{SEL}}-\texttt{SEL}}_\infty
        \le\frac{\gamma}{2\Lambda}\norm{H}_2^2.
    \]
    Choose unitaries \((\texttt{PREP}_L,\texttt{PREP}_R)\) such that
    \begin{align*}
        d_j&=\bra{j}\texttt{PREP}_R\ket{0^{\lceil\log{(k + 1)}\rceil}}=\sqrt{\frac{\abs{w_j}}{\Lambda}}\\
        c_j&=\bra{j}\texttt{PREP}_L\ket{0^{\lceil\log{(k + 1)}\rceil}}=-i\operatorname{sgn}(w_j)d_j
    \end{align*}
    for all \(0 \le j \le k\) and \(c_j^*d_j = 0\) otherwise. Apply \(\texttt{PREP}_R \otimes I\), then the circuit \(U_k \cdots U_1\), and then \(\texttt{PREP}_L^\dagger \otimes I\). Writing the zero block of this operation as \(\frac{\widetilde H}{\Lambda}\), we have
    \begin{align*}
        \norm*{\frac{\widetilde H}{\Lambda}-\frac{B}{\Lambda}}_\infty
        \le
        \norm*{\widetilde{\texttt{SEL}}-\texttt{SEL}}_\infty
    \end{align*}
    which in turn gives
    \begin{align*}
        \norm{\widetilde H-B}_\infty
        \le\Lambda
        \norm{\widetilde{\texttt{SEL}}-\texttt{SEL}}_\infty
        \le\frac{\gamma}{2}\norm{H}_2^2.
    \end{align*}
    From the fact that \(\norm{B-H}_\infty\le\frac{\gamma}{2}\), and \(\norm{B-H}_2\le\left(\frac{\gamma}{2}\right)\norm{H}_2\), and since \(\norm{\cdot}_2\le\norm{\cdot}_\infty\) and \(\norm{H}_2\le1\), we have
    \begin{equation*}
        \norm{\widetilde H-H}_\infty
        \le\frac\gamma2+\frac\gamma2\norm{H}_2^2\le\gamma
    \end{equation*}
    \begin{equation}\label{eq:tilde-H-error-bound}
        \norm{\widetilde H-H}_2
        \le\frac\gamma2\norm{H}_2+\frac\gamma2\norm{H}_2^2
        \le\gamma\norm{H}_2.
    \end{equation}

    The number of queries made to the time evolution operator is \(\bO{k+ \frac{k^3\Lambda}{\gamma}}=\bO{\frac{\log^5\left(\frac{2}{\gamma}\right)}{\gamma}}\), and the total evolution time is \(\sum_{j=1}^kt_j\le k=\bO{\log\left(\frac{2}{\gamma}\right)}\).
\end{proof}

Finally, with this construction of a block encoding of \(H\), we can now easily sample from the Pauli spectrum of \(H\) with any desired precision. Now, we show the following result.

\spectrumsampling*

\begin{proof}
    Using~\cref{lem:block-encoding-H} with accuracy \(\frac{\gamma}{2}\), obtain an \((\alpha, a, \frac{\gamma}{2})\)-block encoding of \(H\), with \(\alpha\) and \(a\) as specified in~\cref{lem:block-encoding-H}. Let \(B\) be the resulting unitary. Its zero block, say \(\frac{\widetilde H}{\alpha}\), satisfies \(\norm{\widetilde H-H}_2\le\frac{\gamma}{2}\norm{H}_2\). By the reverse triangle inequality,
    \[
        \norm{\widetilde H}_2
        \ge\norm{H}_2-\norm{\widetilde H-H}_2
        \ge\frac12\norm{H}_2>0.
    \]

    Consider a system with the following registers: an \(n\)-qubit work register \(\Ac\), another \(n\)-qubit work register \(\Bc\), and an \(a\)-qubit ancilla register \(\Cc\).

    Repeat the following until success:
    \begin{itemize}
        \item Prepare \(\ket\Phi\) in \(\Ac\Bc\) and \(\ket{0^a}\) in \(\Cc\).
        \item Apply \(B\) to \(\Cc\Bc\), and measure the register \(\Cc\).
        \item If the outcome of the measurement is \(\ket{0^a}\), return the state in registers \(\Ac\Bc\). Otherwise, repeat.
    \end{itemize}
    A successful attempt leaves the unnormalized state \(\frac{(I\otimes\widetilde H)\ket\Phi}{\alpha}\). Its normalization gives \(\ket{\Psi(\widetilde H)}\). The probability of successfully obtaining \(\ket{\Psi(\widetilde H)}\) is
    \[
        p=\frac{\norm{\widetilde H}_2^2}{\alpha^2}
        \ge\frac{\norm{H}_2^2}{4\alpha^2}.
    \]
    Therefore, the expected number of attempts is \(p^{-1}\le\frac{4\alpha^2}{\norm{H}_2^2}\). The probability of \(R\) consecutive failures is at most \((1-p)^R\le e^{-pR}\). Hence, with probability at least \(1-\delta\), at least one of the first
    \[
        R=\ceil*{\frac{4\alpha^2}{\norm{H}_2^2}\log\frac1\delta}
    \]
    attempts is successful. Conditioned on success, the returned state is always \(\ket{\Psi(\widetilde H)}\).

    We then measure the state \(\ket{\Psi(\widetilde H)}\) in the Bell basis, giving
    \[
        \prob[Q\sim\wh\Dc]{P=Q}
        =\frac{\abs{\widehat{\widetilde H}(P)}^2}
        {\norm{\widetilde H}_2^2}.
    \]
    Let \(u_P=\frac{\widehat{\widetilde H}(P)}{\norm{\widetilde H}_2}\) and \(v_P=\frac{\widehat H(P)}{\norm{H}_2}\). Parseval's theorem gives \(\sum_P\abs{u_P}^2=\sum_P\abs{v_P}^2=1\). Factoring the difference of squared magnitudes and applying Cauchy--Schwarz, we obtain
    \begin{align*}
        \norm{\wh\Dc-\Dc}_{\mathrm{TV}}
        &=\frac12\sum_P\abs*{\abs{u_P}^2-\abs{v_P}^2}\\
        &\le\frac12
        \left(\sum_P\abs{u_P-v_P}^2\right)^{1/2}
        \left(\sum_P(\abs{u_P}+\abs{v_P})^2\right)^{1/2}\\
        &\le\norm*{
        \frac{\widetilde H}{\norm{\widetilde H}_2}
        -\frac{H}{\norm{H}_2}}_2.
    \end{align*}
    Finally, inserting \(\frac{\widetilde H}{\norm{H}_2}\) gives
    \begin{align*}
        \norm*{
        \frac{\widetilde H}{\norm{\widetilde H}_2}
        -\frac{H}{\norm{H}_2}}_2
        &\le
        \norm*{\frac{\widetilde H}{\norm{\widetilde H}_2}
        -\frac{\widetilde H}{\norm{H}_2}}_2
        +\frac{\norm{\widetilde H-H}_2}{\norm{H}_2}\\
        &=\frac{\abs{\norm{H}_2-\norm{\widetilde H}_2}
        +\norm{\widetilde H-H}_2}{\norm{H}_2}\\
        &\le\frac{2\norm{\widetilde H-H}_2}{\norm{H}_2}
        \le\gamma.
    \end{align*}

    Each attempt at preparing \(\ket{\Psi(\widetilde H)}\) uses \(\bO{\frac{\log^5\left(\frac{2}{\gamma}\right)}{\gamma}}\) queries to the time evolution operator and \(\bO{\log\left(\frac{2}{\gamma}\right)}\) total evolution time. These costs, and the fact that \(\alpha = \bO{\log^2\left(\frac{2}{\gamma}\right)}\), gives us the claimed costs.
\end{proof}

\section{Property Testing}
\label{sec:property-testing}

Let \(H\) be a Hamiltonian, \(\Pc\) be a property (some subset of all Hermitian matrices). A Hamiltonian \(H\) is \(\eps\)-far from \(\Pc\) if \(\norm{H - H'}_2 \ge \eps\) for all \(H' \in \Pc\); \(H\) is \(\eps\)-close to \(\Pc\) if \(\norm{H - H'}_2 \le \eps\) for some \(H' \in \Pc\). The Hamiltonian property testing problem is as follows.

\begin{definition}[Tolerant Hamiltonian property testing]
    Given \(0 \le \eps_1 < \eps_2 < 1\), the \emph{tolerant Hamiltonian property testing} problem is to distinguish between the following two cases
    \begin{enumerate}[label=(\roman*)]
        \item \(H\) is \(\eps_1\)-close to \(\Pc\)
        \item \(H\) is \(\eps_2\)-far from \(\Pc\)
    \end{enumerate}

    An \((\eps_1, \eps_2, \delta)\)-property tester (sometimes abbreviated to \((\eps_1, \eps_2, \delta)\)-tester) for \(\Pc\) is an algorithm that receives as input either an \(H\) that is \(\eps_1\)-close to \(\Pc\) or an \(H\) that is \(\eps_2\)-far from \(\Pc\). In the former case, the algorithm accepts with probability at least \(1 - \delta\); in the latter case, the algorithm rejects with probability at least \(1 - \delta\). If the Hamiltonian satisfies neither side of the corresponding promise, then either output is allowed.
\end{definition}
These definitions can be naturally extended to properties defined on tuples of Hamiltonians \((H_1, \dots, H_m)\). For any property \(\Pc\), the standard Hamiltonian property testing problem is a special case of the tolerant Hamiltonian property testing problem, setting \(\eps_1 = 0\) and \(\eps_2 = \eps\). We will extensively use the term \emph{tail weight} which we define as follows.
\begin{definition}[Tail weight]
    For any Hamiltonian \(H = \sum_{P \in \Pbb_n} \alpha_P P\) and set \(S\subseteq \Pbb_n\), let \(T_H(S) \coloneqq \sum_{P \in \Pbb_n \setminus S} \abs{\alpha_P}^2\) be the tail weight of \(S\).
\end{definition}
When $H$ is unambiguous from the context we will drop the subscript and write $T(S)$.

\subsection{Sparsity and Degree-Boundedness Tester}

In this subsection, we give algorithms for sparsity testing and degree-boundedness testing. We first define sparsity and degree-boundedness. 

\begin{definition}[$s$-sparsity]
A (traceless) Hamiltonian \(H\) is said to be \(s\)-sparse if there exists a set of Paulis \(T \subseteq \Pbb_n\) of size at most \(s\), such that every Pauli in the support of \(H\) belongs to \(T\).
\end{definition}

The degree of a qubit \(j\), \(\deg_j(H)\), is the number of non-zero Pauli terms that act non-trivially on it. Therefore
\begin{align*}
    \deg_j(H)
    &=\abs*{\set{P\in\Pbb_n:
        \wh{H}(P)\neq0,\ P_j\neq I}},\\
    \deg(H)&=\max_{j\in[n]}\deg_j(H).
\end{align*}
where \(\deg(H)\) is the degree of the Hamiltonian.

\begin{definition}[$\Delta$-degree-boundedness]
    A Hamiltonian \(H\) is said to be \(\Delta\)-degree-bounded if \(\deg(H) \le \Delta\).
\end{definition}

We treat both as instances of a more general problem we call \emph{generalized cardinality testing}.

\begin{definition}[Generalized Cardinality]
\label{dfn:gencar}
    Let \(\Sc\) be a set of subsets of \(\Pbb_n\) and \(\ell: \Sbb \rightarrow \Zbb_{\ge 0}\) a function that satisfies the following:
    \begin{enumerate}
        \item There exists a unique \(S_0 \in \Sc\) such that \(\ell(S_0) = 0\). For all \(S \in \Sc, S_0 \subseteq S\).
        \item For all \(S, S' \in \Sc\) such that \(S \subseteq S', \ell(S) \le \ell(S')\).
        \item There exists a fixed positive integer \(b\), such that for all \(S \in \Sc, \abs{S} \le b \ell(S)\). 
    \end{enumerate}

    For any such \((\Sc, \ell)\), we say that \(H\) has \((\Sc, \ell)\)-cardinality \(s\) if there is an \(S \in \Sc\) with \(\ell(S) \le s\) that contains the support of \(H\) in the Pauli basis.
\end{definition}
Note that we define \((\Sc, \ell)\)-cardinality in terms of an upper bound, so every $H$ with cardinality $k$ also has cardinality $k+1$. This is for compatibility with standard property testing definitions.

The principal result of this subsection is an algorithm for the generalized cardinality testing problem.
\begin{restatable}{theorem}
{SetPropertyTester}
\label{thm:set-property-tester}
    Let \(0 \le \eps_1 < \eps_2 < 1\), \(\delta \in (0,1/2)\) and \(s \in \Zbb_{\geq 1}\), and let \((\Sc,\ell)\) be an instance of the generalized cardinality problem.

    There is an $(\eps_1, \eps_2,\delta)$-tester for the property of a Hamiltonian $H$ having $(\Sc,\ell)$-cardinality at most $s$. The tester uses 
    \[
        \bO{\frac{sb}{(\eps_2-\eps_1)^2}
        \log^2\left(\frac{sb}{\delta(\eps_2-\eps_1)}\right)
        \log^5\left(\frac{2}{\eps_2-\eps_1}\right)}
    \]
    total Hamiltonian evolution time and
    \[
        \bO{\frac{sb}{\eps_2(\eps_2-\eps_1)^3}
        \log^2\left(\frac{sb}{\delta(\eps_2-\eps_1)}\right)
        \log^9\left(\frac{2}{\eps_2-\eps_1}\right)}
    \]
    queries to the time evolution operator.
\end{restatable}

We now describe the algorithm:
\begin{enumerate}
    \item Using~\cref{lem:norm-estimate}, with probability at least \(1 - \frac{\delta}{3}\), we either learn that \(\norm{H}_2^2 < \eps_2^2\), or we obtain a \(\left(1 \pm \frac{\eps_2^2 - \eps_1^2}{100\eps_2^2}\right)\)-approximation of \(\norm{H}_2^2\). If \(\norm{H}_2^2 < \eps_2^2\), accept. Otherwise, continue.
    
    \item We maintain a set \(C\) of Paulis, which is initially empty. Then, for \(N_1\) iterations, we sample a Pauli \(P\) from the Pauli spectrum of \(H\) using the Pauli sampling procedure of~\cref{thm:spectrum-sampling-H} and set \(C\) to \(C \cup \{P\}\). \(N_1\) is chosen such that, with probability at least \(1 - \frac{\delta}{3}\), for any \(S \in \Sc\) with \(\ell(S) \le s\), \[\sum_{P \in S \setminus C} \abs{\alpha_P}^2 \le \frac{\eps_2^2 - \eps_1^2}{10}.\]
    
    \item We obtain \(N_2\) samples from the Pauli spectrum of \(H\) using the Pauli sampling procedure of~\cref{thm:spectrum-sampling-H}. \(N_2\) is chosen such that, with probability at least \(1 - \frac{\delta}{3}\), for every \(S \in \Sc\) such that \(\ell(S) \le s\) we either conclude that \(T(S\cap C) \ge 5\eps_2^2\), or estimate \(T(S \cap C)\) to additive error \(\frac{\eps_2^2 - \eps_1^2}{10}\). The algorithm accepts iff there is some \(S \in \Sc\) such that \(\ell(S) \le s\) for which our estimate of \(T(S \cap C)\) is at most
    \[\frac{\eps_1^2 + \eps_2^2}{2}.\]
\end{enumerate}

We first prove that we can collect a set \(C\) as described in step \(2\) of the above algorithm.

\begin{lemma}\label{lem:cardinality-candidates}
    Let \(0 \le \eps_1 < \eps_2 < 1\), \(\delta \in (0,1/2)\) and \(s \in \Zbb_{\ge 1}\), and let \((\Sc,\ell)\) be an instance of the generalized cardinality problem. Suppose we are given a \(\left(1 \pm \frac{\eps_2^2-\eps_1^2}{100\eps_2^2}\right)\)-approximation \(h\) of \(\norm{H}_2^2\), with \(h \ge \frac{\eps_2^2}{2}\).

    There is an algorithm that, with probability at least \(1-\delta\), obtains a set \(C\) of Paulis such that, for every \(S \in \Sc\) with \(\ell(S) \le s\),
    \(
        \sum_{P \in S \setminus C} \abs{\alpha_P}^2 \le \frac{\eps_2^2-\eps_1^2}{10}.
    \)
    The algorithm uses
    \[
        \bO{\frac{sb}{\eps_2(\eps_2-\eps_1)}
        \log^2{\left(\frac{sb}{\delta(\eps_2 - \eps_1)}\right)}
        \log^5\left(\frac{2}{\eps_2 - \eps_1}\right)}
    \]
    total Hamiltonian evolution time and
    \[
        \bO{\frac{sb}{\eps_2^2(\eps_2-\eps_1)^2}
        \log^2{\left(\frac{sb}{\delta(\eps_2 - \eps_1)}\right)}
        \log^9\left(\frac{2}{\eps_2 - \eps_1}\right)}
    \]
    queries to the time evolution operator.
\end{lemma}

\begin{proof}
    Set 
    \[
        N_1=\ceil*{
        \frac{50sbh}{\eps_2^2-\eps_1^2}
        \log\left(\frac{12sbh}{\delta(\eps_2^2-\eps_1^2)}\right)}
    \]
    Set \(\gamma=\frac{\eps_2^2-\eps_1^2}{100}\) to be the accuracy and \(\frac{\delta}{2N_1}\) to be the failure probability with which we sample from the Pauli spectrum using~\cref{thm:spectrum-sampling-H}. Obtain \(N_1\) independent samples, and let \(C\) be the set of Paulis that appear. A union bound bounds the probability of any sampling failure by \(\frac\delta2\). The sampling failure probability bounds the chance of exceeding the stated costs of sampling.

    Let \(\widetilde\Dc\) be the sampling distribution, and let \(\widetilde p_P\) be the probability of sampling \(P\). Every Pauli \(P\) with
    \[
        \widetilde p_P \ge \frac{\eps_2^2-\eps_1^2}{20sbh}
    \]
    is missed with probability at most
    \[
        \exp\left(-\frac{N_1(\eps_2^2-\eps_1^2)}{20sbh}\right).
    \]
    There are at most \(\frac{20sbh}{\eps_2^2-\eps_1^2}\) such Paulis by Parseval's theorem. The choice of \(N_1\), followed by a union bound, therefore ensures that all such Paulis belong to \(C\) with probability at least \(1-\frac\delta2\).

    On this event, since \(\abs{S}\le sb\) for every \(S\in\Sc\) with \(\ell(S)\le s\),
    \[
        \sum_{P\in S\setminus C}\widetilde p_P
        \le \frac{\eps_2^2-\eps_1^2}{20h}.
    \]
    By the total variation distance guarantee of~\cref{thm:spectrum-sampling-H},
    \begin{align*}
        \sum_{P\in S\setminus C}\abs{\alpha_P}^2
        &\le \norm{H}_2^2
        \left(\sum_{P\in S\setminus C}\widetilde p_P+\gamma\right)\\
        &\le \frac{\norm{H}_2^2}{20h}(\eps_2^2-\eps_1^2)
        +\norm{H}_2^2\gamma\\
        &\le \frac{\eps_2^2-\eps_1^2}{10},
    \end{align*}
    where we used \(h\ge\frac{99}{100}\norm{H}_2^2\) and \(\norm{H}_2^2\le1\). A union bound gives the required success probability.

    Since \(h=\Theta(\norm{H}_2^2)\),~\cref{thm:spectrum-sampling-H} gives total evolution time \(\bO{\frac{N_1}{\norm{H}_2^2}\log^5\left(\frac{2}{\gamma}\right)}\) and query complexity \(\bO{\frac{N_1}{\norm{H}_2^2\gamma}\log^9\left(\frac{2}{\gamma}\right)}\). This, combined with \(\eps_2 (\eps_2 - \eps_1) \le \eps_2^2 - \eps_1^2 \le 2 \eps_2 (\eps_2 - \eps_1)\) gives us the stated bounds.
\end{proof}

We now prove that we can, for each \(S \in \Sc\) such that \(\ell(S) \le s\), either conclude that \(T(S\cap C)\) is large, or estimate it to desired accuracy.

\begin{lemma}\label{lem:cardinality-tail-estimation}
    Let \(0 \le \eps_1 < \eps_2 < 1\), \(\delta \in (0,1/2)\) and \(s \in \Zbb_{\ge 1}\), and let \((\Sc,\ell)\) be an instance of the generalized cardinality problem. Suppose we are given a \(\left(1 \pm \frac{\eps_2^2-\eps_1^2}{100\eps_2^2}\right)\)-approximation \(h\) of \(\norm{H}_2^2\), with \(h \ge \frac{\eps_2^2}{2}\), and a set \(C\) of at most \(N_1\) Paulis, where \(N_1\ge1\).

    There is an algorithm that, with probability at least \(1-\delta\), for every \(S\in\Sc\) with \(\ell(S)\le s\), either certifies that \(T(S\cap C)\ge5\eps_2^2\), or estimates \(T(S\cap C)\) to additive error at most \(\frac{\eps_2^2-\eps_1^2}{10}\). The algorithm uses
    \[
        \bO{\frac{sb}{(\eps_2-\eps_1)^2}
        \log{\left(\frac{N_1+1}{\delta}\right)}
        \log\left(\frac{sb(N_1+1)}{\delta(\eps_2-\eps_1)}\right)
        \log^5\left(\frac{2}{\eps_2-\eps_1}\right)}
    \]
    total Hamiltonian evolution time and
    \[
        \bO{\frac{sb}{\eps_2(\eps_2-\eps_1)^3}
        \log{\left(\frac{N_1+1}{\delta}\right)}
        \log\left(\frac{sb(N_1+1)}{\delta(\eps_2-\eps_1)}\right)
        \log^9\left(\frac{2}{\eps_2-\eps_1}\right)}
    \]
    queries to the time evolution operator.
\end{lemma}

\begin{proof}
    Consider the sets \(S\cap C\), where \(S\in\Sc\) and \(\ell(S)\le s\). Each set has cardinality at most \(sb\), so the number of distinct such sets is at most \((N_1+1)^{sb}\).

    Set
    \[
        N_2=\ceil*{
        \frac{200000sb\eps_2^2h}{(\eps_2^2-\eps_1^2)^2}
        \log\left(\frac{12(N_1+1)}{\delta}\right)}.
    \]
    Obtain \(N_2\) fresh independent samples using~\cref{thm:spectrum-sampling-H}, with accuracy \(\gamma=\frac{\eps_2^2-\eps_1^2}{100}\) and failure probability \(\frac{\delta}{2N_2}\) for each sample. A union bound bounds the probability of any sampling failure by \(\frac\delta2\). The sampling failure probability bounds the chance of exceeding the stated costs of sampling.

    For each set \(S\cap C\), let \(\widetilde p(S\cap C)\) be the probability that a sample lies outside \(S\cap C\), and let \(\widehat p(S\cap C)\) be the fraction of samples that lie outside \(S\cap C\). Set
    \[
        \widehat T(S\cap C)=h\widehat p(S\cap C).
    \]
    If \(\widehat T(S\cap C)\ge6\eps_2^2\), certify that \(T(S\cap C)\ge5\eps_2^2\). Otherwise, return \(\widehat T(S\cap C)\) as the estimate.

    By the total variation distance guarantee of~\cref{thm:spectrum-sampling-H},
    \[
        \abs*{\widetilde p(S\cap C)
        -\frac{T(S\cap C)}{\norm{H}_2^2}}
        \le\gamma.
    \]
    Consequently,
    \[
        \abs{h\widetilde p(S\cap C)-T(S\cap C)}
        \le \frac{\eps_2^2-\eps_1^2}{100\eps_2^2}T(S\cap C)
        +h\gamma.
    \]
    In particular, if \(T(S\cap C)<5\eps_2^2\), then \(h\widetilde p(S\cap C)<5.1\eps_2^2\), and if \(T(S\cap C)>7\eps_2^2\), then \(h\widetilde p(S\cap C)>6.9\eps_2^2\). By~\cref{lem:chernoff-prelim}, with probability at least \(1-\frac{\delta}{4(N_1+1)^{sb}}\), the certificate is correct whenever it is returned, and otherwise \(T(S\cap C)\le7\eps_2^2\).

    For a set with \(T(S\cap C)\le7\eps_2^2\), we have
    \[
        \widetilde p(S\cap C)
        \le\frac{7\eps_2^2}{\norm{H}_2^2}+\gamma.
    \]
    Therefore by~\cref{lem:bernstein-prelim} and the choice of \(N_2\), with probability at least \(1-\frac{\delta}{4(N_1+1)^{sb}}\),
    \[
        h\abs{\widehat p(S\cap C)-\widetilde p(S\cap C)}
        \le\frac{\eps_2^2-\eps_1^2}{100}.
    \]
    Since \(h\le\frac{101}{100}\norm{H}_2^2\le\frac{101}{100}\), this gives
    \begin{align*}
        \abs{\widehat T(S\cap C)-T(S\cap C)}
        &\le h\abs{\widehat p(S\cap C)-\widetilde p(S\cap C)}
        +\abs{h\widetilde p(S\cap C)-T(S\cap C)}\\
        &\le \frac{\eps_2^2-\eps_1^2}{100}
        +\frac{7(\eps_2^2-\eps_1^2)}{100}
        +\frac{101(\eps_2^2-\eps_1^2)}{10000}\\
        &<\frac{\eps_2^2-\eps_1^2}{10}.
    \end{align*}
    A union bound over the at most \((N_1+1)^{sb}\) sets gives simultaneous correctness with probability at least \(1-\frac\delta2\). Another union bound with the sampling failures gives the required success probability.

    Since \(h=\Theta(\norm{H}_2^2)\),~\cref{thm:spectrum-sampling-H} gives total evolution time \(\bO{\frac{N_2}{\norm{H}_2^2}\log\left(\frac{2N_2}{\delta}\right)\log^5\left(\frac{2}{\gamma}\right)}\) and query complexity \(\bO{\frac{N_2}{\norm{H}_2^2\gamma}\log\left(\frac{2N_2}{\delta}\right)\log^9\left(\frac{2}{\gamma}\right)}\). This, combined with \(\eps_2 (\eps_2 - \eps_1) \le \eps_2^2 - \eps_1^2 \le 2 \eps_2 (\eps_2 - \eps_1)\) gives us the stated bounds.
\end{proof}

We are now ready to prove our theorem.

\begin{proof}[Proof of~\cref{thm:set-property-tester}]
    Use~\cref{lem:norm-estimate} with failure probability \(\frac\delta3\). If it certifies that \(\norm{H}_2^2<\eps_2^2\), accept, since the zero Hamiltonian has \((\Sc,\ell)\)-cardinality at most \(s\). Otherwise, let \(h\) be the returned \(\left(1\pm\frac{\eps_2^2-\eps_1^2}{100\eps_2^2}\right)\)-approximation of \(\norm{H}_2^2\).

    Apply~\cref{lem:cardinality-candidates} to obtain \(C\), and then~\cref{lem:cardinality-tail-estimation} to its sets, each with failure probability \(\frac\delta3\). Condition on all three stages succeeding, which occurs with probability at least \(1-\delta\).

    For completeness, suppose \(H\) is \(\eps_1\)-close to having \((\Sc,\ell)\)-cardinality at most \(s\). By Parseval's theorem, there is an \(S_*\in\Sc\), with \(\ell(S_*)\le s\), such that \(T(S_*)\le\eps_1^2\). By~\cref{lem:cardinality-candidates},
    \[
        T(S_*\cap C)
        =T(S_*)+\sum_{P\in S_*\setminus C}\abs{\alpha_P}^2
        \le\eps_1^2+\frac{\eps_2^2-\eps_1^2}{10}.
    \]
    This set is not certified large, and its estimate is at most
    \[
        \eps_1^2+\frac{\eps_2^2-\eps_1^2}{5}
        <\frac{\eps_1^2+\eps_2^2}{2}.
    \]
    Hence the tester accepts.

    For soundness, suppose \(H\) is \(\eps_2\)-far from having \((\Sc,\ell)\)-cardinality at most \(s\). Every set considered is contained in an \(S \in \Sc\) with \(\ell(S) \le s\), so Parseval's theorem implies that its tail weight is at least \(\eps_2^2\). It is therefore either certified large or has estimate at least
    \[
        \eps_2^2-\frac{\eps_2^2-\eps_1^2}{10}
        >\frac{\eps_1^2+\eps_2^2}{2}.
    \]
    Thus the tester rejects.

    The costs in~\cref{lem:cardinality-tail-estimation} subsume those in~\cref{lem:cardinality-candidates,lem:norm-estimate}. Substituting the bound on \(N_1\), the total evolution time is
    \[
        \bO{\frac{sb}{(\eps_2-\eps_1)^2}
        \log^2\left(\frac{sb}{\delta(\eps_2-\eps_1)}\right)
        \log^5\left(\frac{2}{\eps_2-\eps_1}\right)}
    \]
    and the number of queries to the time evolution operator is
    \[
        \bO{\frac{sb}{\eps_2(\eps_2-\eps_1)^3}
        \log^2\left(\frac{sb}{\delta(\eps_2-\eps_1)}\right)
        \log^9\left(\frac{2}{\eps_2-\eps_1}\right)}.
    \]
\end{proof}

We first apply~\cref{thm:set-property-tester} to the sparsity testing problem.

\begin{restatable}{theorem}{sparsity-tolerant}\label{thm:sparsity-tester-tolerant}
    For all \(0 \le \eps_1 < \eps_2 < 1\), \(\delta \in (0,1/2)\), there is an $(\eps_1, \eps_2,\delta)$-tester for having sparsity \(s \in \Zbb_{\ge 1}\) that uses 
    \[
        \bO{\frac{s}{(\eps_2-\eps_1)^2}
        \log^2\left(\frac{s}{\delta(\eps_2-\eps_1)}\right)
        \log^5\left(\frac{2}{\eps_2-\eps_1}\right)}
    \]
    total Hamiltonian evolution time and
    \[
        \bO{\frac{s}{\eps_2(\eps_2-\eps_1)^3}
        \log^2\left(\frac{s}{\delta(\eps_2-\eps_1)}\right)
        \log^9\left(\frac{2}{\eps_2-\eps_1}\right)}
    \]
    queries to the time evolution operator.
\end{restatable}

\begin{proof}[Proof of~\cref{thm:sparsity-tester-tolerant}]
    Let \(\Sc\) be the power set of \(\Pbb_n\) and \(\ell(S) = \abs{S}\). Observe that \(b = 1\) for the chosen \(\ell\). This satisfies the requirements of~\cref{dfn:gencar}. \(H\) is \(s\)-sparse iff \(\ell(\supp(H)) \le s\), so we may apply~\cref{thm:set-property-tester}.
\end{proof}

We now consider the degree-boundedness testing problem.

\begin{restatable}{theorem}{degreeboundedness-tolerant}\label{thm:degree-boundedness-tester-tolerant}
    For all \(0 \le \eps_1 < \eps_2 < 1\), \(\delta \in (0,1/2)\), there is an $(\eps_1, \eps_2,\delta)$-tester for having degree-boundedness \(\Delta  \in \Zbb_{\ge 1}\) that uses 
    \[
        \bO{\frac{n\Delta}{(\eps_2-\eps_1)^2}
        \log^2\left(\frac{n\Delta}{\delta(\eps_2-\eps_1)}\right)
        \log^5\left(\frac{2}{\eps_2-\eps_1}\right)}
    \]
    total Hamiltonian evolution time and
    \[
        \bO{\frac{n\Delta}{\eps_2(\eps_2-\eps_1)^3}
        \log^2\left(\frac{n\Delta}{\delta(\eps_2-\eps_1)}\right)
        \log^9\left(\frac{2}{\eps_2-\eps_1}\right)}
    \]
    queries to the time evolution operator.
\end{restatable}

\begin{proof}[Proof of~\cref{thm:degree-boundedness-tester-tolerant}]
    Let \(\Sc\) be the power set of \(\Pbb_n \setminus \{I^{\otimes n}\}\) and \(\ell(S) = \\ \max_{j\in[n]}{\abs{\left\{ P \in S: P_j \neq I \right\}}}\). Observe that \(b = n\) for the chosen \(\ell\). This satisfies the requirements of~\cref{dfn:gencar}. \(H\) is \(\Delta\)-degree-bounded iff \(\ell(\supp(H)) \le \Delta\), so we may apply~\cref{thm:set-property-tester}.
\end{proof}

\subsection{Junta and Pauli Dimensionality Tester}

In this subsection, we give algorithms for junta testing and Pauli dimensionality testing. We first define juntaness and Pauli dimensionality. 

\begin{definition}[$k$-juntas]
    A Hamiltonian \(H\) is a \(k\)-junta if there exists \( J \subseteq [n],
    \abs{J} \le k\) such that, for all Paulis \(P \in \supp(H)\), $P_i = I$ for
    every $i \notin J$.
\end{definition}

\begin{definition}[$k$-Pauli dimensionality]
    A Hamiltonian \(H\) has Pauli dimensionality $k$ if its support is
    contained in a subgroup of $\Pbb_n$ with a generating set of size at most
    $k$.
\end{definition}

We treat both as instances of a more general problem we call \emph{generalized dimensionality testing}. Let \(\{\pm I, \pm X, \pm Y, \pm Z, \pm i I, \pm i X, \pm i Y, \pm i Z\}^{\otimes n}\) be the Pauli group on \(n\) qubits. \\ Therefore \(\Pbb_n = \{\pm I, \pm X, \pm Y, \pm Z, \pm i I, \pm i X, \pm i Y, \pm i Z\}^{\otimes n} / \{\pm 1, \pm i\}\) is the quotient group of the Pauli group. 

\begin{definition}[Generalized Dimensionality]
\label{dfn:gendim}
Let \(\Sc\) be a set of subgroups of \(\Pbb_n\) and \(\ell: \Sc \rightarrow \Zbb_{\geq 0}\) a function that satisfies the following:
\begin{enumerate}
    \item There exists a unique \(G_0 \in \Sc\) such that \(\ell(G_0) = 0\).  For all \(G \in \Sc\), \(G_0 \subseteq G\).  
    \item For all \(G \in \Sc, P \in \Pbb_n \setminus G\), let $\Sc_{G,P} = \set{G' \in \Sc : G \subseteq G', P \in G'}$. There is a unique ``minimal'' $G'' \in \Sc_{G,P}$ that is contained in every other element of $\Sc_{G,P}$.  Furthermore, $\ell(G'') \ge \ell(G) + 1$.
\end{enumerate}

For any such $(\Sc,\ell)$, we say that $H$ has $(\Sc,\ell)$-dimensionality $k$ if there is a $G \in \Sc$ with $\ell(G) \le k$ that contains the support of $H$ in the Pauli basis.
\end{definition}
Just like \((\Sc, \ell)\)-cardinality, we define $(\Sc,\ell)$-dimensionality in terms of an upper bound, so
every $H$ with dimensionality $k$ also has dimensionality $k+1$.

The principal result of this subsection is an algorithm for the generalized dimensionality testing problem.
\begin{restatable}{theorem}
{GroupPropertyTester}
\label{thm:group-property-tester}
    Let \(0 \le \eps_1 < \eps_2 < 1\), \(\delta \in (0,1/2)\) and \(k \in \Zbb_{\geq 1}\), and let \((\Sc,\ell)\) be an instance of the generalized dimensionality problem.

    There is an $(\eps_1, \eps_2, \delta)$-tester for the property of a Hamiltonian $H$ having $(\Sc, \ell)$-dimensionality at most $k$.  The tester uses 
    \[
        \bO{\frac{k}{(\eps_2-\eps_1)^2}
        \log^2\left(\frac{k}{\delta(\eps_2-\eps_1)}\right)
        \log^5\left(\frac{2}{\eps_2-\eps_1}\right)}
    \]
    total Hamiltonian evolution time and
    \[
        \bO{\frac{k}{\eps_2(\eps_2-\eps_1)^3}
        \log^2\left(\frac{k}{\delta(\eps_2-\eps_1)}\right)
        \log^9\left(\frac{2}{\eps_2-\eps_1}\right)}
    \]
    queries to the time evolution operator.
\end{restatable}

For any \(U\subseteq \Pbb_n\), let \(G_U\) be the unique minimal member of \(\Sc\) containing \(U\). This is obtained by starting with \(G_0\) and repeatedly applying the last requirement of~\cref{dfn:gendim}.

We now describe our algorithm: 
\begin{enumerate}
    \item Using~\cref{lem:norm-estimate}, with probability at least \(1 - \frac{\delta}{3}\), we either learn that \(\norm{H}_2^2 < \eps_2^2\), or we obtain a \(\left(1 \pm \frac{\eps_2^2 - \eps_1^2}{100\eps_2^2}\right)\)-approximation of \(\norm{H}_2^2\). If \(\norm{H}_2^2 < \eps_2^2\), accept. Otherwise, continue.
    
    \item We maintain a set \(C\) of Paulis, which is initially empty. Then, for \(N_1\) iterations, we sample a Pauli \(P\) from the Pauli spectrum of \(H\) using the Pauli sampling procedure of~\cref{thm:spectrum-sampling-H} and set \(C\) to \(C \cup \{P\}\). Let 
    \[
        W \coloneqq \argmin_{G \in \Sc: \ell(G) \le k} \sum_{P \in \Pbb_n \setminus G} \abs{\alpha_P}^2
    \]\(N_1\) is chosen such that, with probability at least \(1 - \frac{\delta}{3}\), there is a set \(U \subseteq C\) satisfying \(\abs{U} \le k\), \(\ell(G_U) \le k\), and 
    \[ \sum_{P\in W\setminus G_U}\abs{\alpha_P}^2 \le\frac{\eps_2^2-\eps_1^2}{10}. \]
    
    \item We obtain \(N_2\) samples from the Pauli spectrum of \(H\) using the Pauli sampling procedure of~\cref{thm:spectrum-sampling-H}. \(N_2\) is chosen such that, with probability at least \(1 - \frac{\delta}{3}\), for every set \(U \subseteq C\) such that \(\abs{U} \le k\) and \(\ell(G_U) \le k\) we either conclude that \(T(G_U) \ge 5\eps_2^2\), or estimate \(T(G_U)\) to additive error \(\frac{\eps_2^2 - \eps_1^2}{10}\). The algorithm accepts iff there is some such \(U\) for which our estimate of \(T(G_U)\) is at most
    \[\frac{\eps_1^2 + \eps_2^2}{2}.\]
\end{enumerate}

We first prove that we can collect a set \(C\) as described in step \(2\) of the above algorithm.

\begin{lemma}\label{lem:dimensionality-candidates}
    Let \(0 \le \eps_1 < \eps_2 < 1\), \(\delta \in (0,1/2)\) and \(k \in \Zbb_{\ge 1}\), and let \((\Sc,\ell)\) be an instance of the generalized dimensionality problem. Suppose we are given a \(\left(1 \pm \frac{\eps_2^2-\eps_1^2}{100\eps_2^2}\right)\)-approximation \(h\) of \(\norm{H}_2^2\), with \(h \ge \frac{\eps_2^2}{2}\). Let \(W\in\Sc\), with \(\ell(W)\le k\), minimize \(T(W)\).

    There is an algorithm that, with probability at least \(1-\delta\), obtains a set \(C\) of Paulis such that there is a \(U\subseteq C\) satisfying \(\abs{U}\le k\), \(G_U\subseteq W\), \(\ell(G_U)\le k\), and
    \[
        \sum_{P\in W\setminus G_U}\abs{\alpha_P}^2
        \le\frac{\eps_2^2-\eps_1^2}{10}.
    \]
    The algorithm uses
    \[
        \bO{\frac{k}{\eps_2(\eps_2-\eps_1)}
        \log{\left(\frac{k}{\delta}\right)}
        \log\left(\frac{k}{\delta(\eps_2-\eps_1)}\right)
        \log^5\left(\frac{2}{\eps_2-\eps_1}\right)}
    \]
    total Hamiltonian evolution time and
    \[
        \bO{\frac{k}{\eps_2^2(\eps_2-\eps_1)^2}
        \log{\left(\frac{k}{\delta}\right)}
        \log\left(\frac{k}{\delta(\eps_2-\eps_1)}\right)
        \log^9\left(\frac{2}{\eps_2-\eps_1}\right)}
    \]
    queries to the time evolution operator.
\end{lemma}

\begin{proof}
    Set
    \[
        N_1=k\ceil*{
        \frac{50h}{\eps_2^2-\eps_1^2}
        \log\left(\frac{12k}{\delta}\right)}
    \]
    Set \(\gamma=\frac{\eps_2^2-\eps_1^2}{100}\) to be the accuracy and \(\frac{\delta}{2N_1}\) to be the failure probability with which we sample from the Pauli spectrum using~\cref{thm:spectrum-sampling-H}. Obtain \(N_1\) independent samples, and let \(C\) be the set of Paulis that appear. A union bound bounds the probability of any sampling failure by \(\frac\delta2\). The sampling failure probability bounds the chance of exceeding the stated costs of sampling.

    Repeatedly applying the last requirement of~\cref{dfn:gendim} shows that, for any \(G,G'\in\Sc\) with \(G\subsetneq G'\), we have \(\ell(G')\ge\ell(G)+1\): starting from \(G\), add elements of \(G'\) outside the current subgroup until it equals \(G'\).

    Divide the \(N_1\) samples into \(k\) blocks of equal size. For \(i\in\{0,\ldots,k\}\), let \(V_i\) be the unique minimal member of \(\Sc\) containing all Paulis lying in \(W\) observed in the first \(i\) blocks. Therefore \(V_0=G_0\), \(V_i\subseteq W\), \(\ell(V_i)\le k\), and \(V_i\subseteq V_{i+1}\) for \(i<k\).

    If
    \[
        \sum_{P\in W\setminus V_i}\abs{\alpha_P}^2
        \ge\frac{\eps_2^2-\eps_1^2}{10},
    \]
    every sample in block $i+1$ lies in \(W\setminus V_i\) with probability at least
    \[
        \frac{\eps_2^2-\eps_1^2}{10\norm{H}_2^2}-\gamma
        \ge\frac{\eps_2^2-\eps_1^2}{20\norm{H}_2^2}.
    \]
    The probability that the entire block misses this set is therefore at most
    \[
        \exp\left(
        -\frac{N_1(\eps_2^2-\eps_1^2)}{20k\norm{H}_2^2}
        \right)
        \le\frac{\delta}{2k}.
    \]
    Otherwise, \(V_i\subsetneq V_{i+1}\), so \(\ell(V_{i+1})\ge\ell(V_i)+1\).

    A union bound over the \(k\) blocks ensures that every block for which the remaining weight is at least \(\frac{\eps_2^2-\eps_1^2}{10}\) gives a strict extension, with probability at least \(1-\frac\delta2\). If the remaining weight ever falls below this threshold, it stays below it. Otherwise, all \(k\) blocks give strict extensions, so \(\ell(V_k)\ge k\). Since \(V_k\subseteq W\) and \(\ell(W)\le k\), this forces \(V_k=W\). Therefore on this event,
    \[
        \sum_{P\in W\setminus V_k}\abs{\alpha_P}^2
        \le\frac{\eps_2^2-\eps_1^2}{10}.
    \]

    Moreover, \(V_k=G_U\) for some \(U\subseteq C\) with \(\abs{U}\le k\). To see this, start with \(U=\emptyset\) and process the observed Paulis lying in \(W\), adding a Pauli to \(U\) only when it lies outside the current \(G_U\). Each addition increases \(\ell(G_U)\) by at least one, so at most \(k\) Paulis are retained. At the end, \(G_U=V_k\), since it contains all the observed Paulis lying in \(W\). A union bound with the sampling failures gives the required success probability.

    Since \(h=\Theta(\norm{H}_2^2)\),~\cref{thm:spectrum-sampling-H} gives total evolution time \(\bO{\frac{N_1}{\norm{H}_2^2}\log\left(\frac{2N_1}{\delta}\right)\log^5\left(\frac{2}{\gamma}\right)}\) and query complexity \(\bO{\frac{N_1}{\norm{H}_2^2\gamma}\log\left(\frac{2N_1}{\delta}\right)\log^9\left(\frac{2}{\gamma}\right)}\). This, combined with \(\eps_2 (\eps_2 - \eps_1) \le \eps_2^2 - \eps_1^2 \le 2 \eps_2 (\eps_2 - \eps_1)\) gives us the stated bounds.
\end{proof}

We now prove that we can, for each \(U \subseteq C\) such that \(\abs{U} \le k\) and \(\ell(G_U) \le k\), either conclude that \(T(G_U)\) is large, or estimate it to desired accuracy, as described in step \(3\) of the above algorithm.

\begin{lemma}\label{lem:dimensionality-tail-estimation}
    Let \(0 \le \eps_1 < \eps_2 < 1\), \(\delta \in (0,1/2)\) and \(k \in \Zbb_{\ge 1}\), and let \((\Sc,\ell)\) be an instance of the generalized dimensionality problem. Suppose we are given a \(\left(1 \pm \frac{\eps_2^2-\eps_1^2}{100\eps_2^2}\right)\)-approximation \(h\) of \(\norm{H}_2^2\), with \(h \ge \frac{\eps_2^2}{2}\), and a set \(C\) of at most \(N_1\) Paulis, where \(N_1\ge1\).

    There is an algorithm that, with probability at least \(1-\delta\), for every set \(U\subseteq C\) with \(\abs{U}\le k\) and \(\ell(G_U)\le k\), either certifies that \(T(G_U)\ge5\eps_2^2\), or estimates \(T(G_U)\) to additive error at most \(\frac{\eps_2^2-\eps_1^2}{10}\). The algorithm uses
    \[
        \bO{\frac{k}{(\eps_2-\eps_1)^2}
        \log{\left(\frac{N_1+1}{\delta}\right)}
        \log\left(\frac{k(N_1+1)}{\delta(\eps_2-\eps_1)}\right)
        \log^5\left(\frac{2}{\eps_2-\eps_1}\right)}
    \]
    total Hamiltonian evolution time and
    \[
        \bO{\frac{k}{\eps_2(\eps_2-\eps_1)^3}
        \log{\left(\frac{N_1+1}{\delta}\right)}
        \log\left(\frac{k(N_1+1)}{\delta(\eps_2-\eps_1)}\right)
        \log^9\left(\frac{2}{\eps_2-\eps_1}\right)}
    \]
    queries to the time evolution operator.
\end{lemma}

\begin{proof}
    Consider the sets \(U\subseteq C\) such that \(\abs{U}\le k\) and \(\ell(G_U)\le k\). The number of distinct such sets is at most \((N_1+1)^k\).

    Set
    \[
        N_2=\ceil*{
        \frac{200000k\eps_2^2h}{(\eps_2^2-\eps_1^2)^2}
        \log\left(\frac{12(N_1+1)}{\delta}\right)}.
    \]
    Obtain \(N_2\) fresh independent samples using~\cref{thm:spectrum-sampling-H}, with accuracy \(\gamma=\frac{\eps_2^2-\eps_1^2}{100}\) and failure probability \(\frac{\delta}{2N_2}\) for each sample. A union bound bounds the probability of any sampling failure by \(\frac\delta2\). The sampling failure probability bounds the chance of exceeding the stated costs of sampling.

    For each set \(G_U\), let \(\widetilde p(G_U)\) be the probability that a sample lies outside \(G_U\), and let \(\widehat p(G_U)\) be the fraction of samples that lie outside \(G_U\). Set
    \[
        \widehat T(G_U)=h\widehat p(G_U).
    \]
    If \(\widehat T(G_U)\ge6\eps_2^2\), certify that \(T(G_U)\ge5\eps_2^2\). Otherwise, return \(\widehat T(G_U)\) as the estimate.

    By the total variation distance guarantee of~\cref{thm:spectrum-sampling-H},
    \[
        \abs*{\widetilde p(G_U)
        -\frac{T(G_U)}{\norm{H}_2^2}}
        \le\gamma.
    \]
    Consequently,
    \[
        \abs{h\widetilde p(G_U)-T(G_U)}
        \le \frac{\eps_2^2-\eps_1^2}{100\eps_2^2}T(G_U)
        +h\gamma.
    \]
    In particular, if \(T(G_U)<5\eps_2^2\), then \(h\widetilde p(G_U)<5.1\eps_2^2\), and if \(T(G_U)>7\eps_2^2\), then \(h\widetilde p(G_U)>6.9\eps_2^2\). By~\cref{lem:chernoff-prelim}, with probability at least \(1-\frac{\delta}{4(N_1+1)^k}\), the certificate is correct whenever it is returned, and otherwise \(T(G_U)\le7\eps_2^2\).

    For a set with \(T(G_U)\le7\eps_2^2\), we have
    \[
        \widetilde p(G_U)
        \le\frac{7\eps_2^2}{\norm{H}_2^2}+\gamma.
    \]
    Therefore by~\cref{lem:bernstein-prelim} and the choice of \(N_2\), with probability at least \(1-\frac{\delta}{4(N_1+1)^k}\),
    \[
        h\abs{\widehat p(G_U)-\widetilde p(G_U)}
        \le\frac{\eps_2^2-\eps_1^2}{100}.
    \]
    Since \(h\le\frac{101}{100}\norm{H}_2^2\le\frac{101}{100}\), this gives
    \begin{align*}
        \abs{\widehat T(G_U)-T(G_U)}
        &\le h\abs{\widehat p(G_U)-\widetilde p(G_U)}
        +\abs{h\widetilde p(G_U)-T(G_U)}\\
        &\le \frac{\eps_2^2-\eps_1^2}{100}
        +\frac{7(\eps_2^2-\eps_1^2)}{100}
        +\frac{101(\eps_2^2-\eps_1^2)}{10000}\\
        &<\frac{\eps_2^2-\eps_1^2}{10}.
    \end{align*}
    A union bound over the at most \((N_1+1)^k\) sets gives simultaneous correctness with probability at least \(1-\frac\delta2\). Another union bound with the sampling failures gives the required success probability.

    Since \(h=\Theta(\norm{H}_2^2)\),~\cref{thm:spectrum-sampling-H} gives total evolution time \(\bO{\frac{N_2}{\norm{H}_2^2}\log\left(\frac{2N_2}{\delta}\right)\log^5\left(\frac{2}{\gamma}\right)}\) and query complexity \(\bO{\frac{N_2}{\norm{H}_2^2\gamma}\log\left(\frac{2N_2}{\delta}\right)\log^9\left(\frac{2}{\gamma}\right)}\). This, combined with \(\eps_2 (\eps_2 - \eps_1) \le \eps_2^2 - \eps_1^2 \le 2 \eps_2 (\eps_2 - \eps_1)\) gives us the stated bounds.
\end{proof}

We are now ready to prove our theorem.

\begin{proof}[Proof of~\cref{thm:group-property-tester}]
    Use~\cref{lem:norm-estimate} with failure probability \(\frac\delta3\). If it certifies that \(\norm{H}_2^2<\eps_2^2\), accept, since the zero Hamiltonian has \((\Sc,\ell)\)-dimensionality at most \(k\). Otherwise, let \(h\) be the returned \(\left(1\pm\frac{\eps_2^2-\eps_1^2}{100\eps_2^2}\right)\)-approximation of \(\norm{H}_2^2\).

    Apply~\cref{lem:dimensionality-candidates} to obtain \(C\), and then~\cref{lem:dimensionality-tail-estimation} to its sets, each with failure probability \(\frac\delta3\). Condition on all three stages succeeding, which occurs with probability at least \(1-\delta\).

    For completeness, suppose \(H\) is \(\eps_1\)-close to having \((\Sc,\ell)\)-dimensionality at most \(k\). By Parseval's theorem, the minimizing subgroup \(W\) satisfies \(T(W)\le\eps_1^2\). By~\cref{lem:dimensionality-candidates}, there is a set \(U\subseteq C\), with \(\abs{U}\le k\), \(G_U\subseteq W\), and \(\ell(G_U)\le k\), such that
    \[
        T(G_U)
        =T(W)+\sum_{P\in W\setminus G_U}\abs{\alpha_P}^2
        \le\eps_1^2+\frac{\eps_2^2-\eps_1^2}{10}.
    \]
    This set is not certified large, and its estimate is at most
    \[
        \eps_1^2+\frac{\eps_2^2-\eps_1^2}{5}
        <\frac{\eps_1^2+\eps_2^2}{2}.
    \]
    Hence the tester accepts.

    For soundness, suppose \(H\) is \(\eps_2\)-far from having \((\Sc,\ell)\)-dimensionality at most \(k\). For every set \(U\) considered, \(G_U\) belongs to \(\Sc\) and has \(\ell(G_U)\le k\), so Parseval's theorem implies that its tail weight is at least \(\eps_2^2\). It is therefore either certified large or has estimate at least
    \[
        \eps_2^2-\frac{\eps_2^2-\eps_1^2}{10}
        >\frac{\eps_1^2+\eps_2^2}{2}.
    \]
    Thus the tester rejects.

    The costs in~\cref{lem:dimensionality-tail-estimation} subsume those in~\cref{lem:dimensionality-candidates,lem:norm-estimate}. Substituting the bound on \(N_1\), the total evolution time is
    \[
        \bO{\frac{k}{(\eps_2-\eps_1)^2}
        \log^2\left(\frac{k}{\delta(\eps_2-\eps_1)}\right)
        \log^5\left(\frac{2}{\eps_2-\eps_1}\right)}
    \]
    and the number of queries to the time evolution operator is
    \[
        \bO{\frac{k}{\eps_2(\eps_2-\eps_1)^3}
        \log^2\left(\frac{k}{\delta(\eps_2-\eps_1)}\right)
        \log^9\left(\frac{2}{\eps_2-\eps_1}\right)}.
    \]
\end{proof}

We first apply~\cref{thm:group-property-tester} to the junta testing problem. 

\begin{restatable}{theorem}{juntaness-tolerant}
  \label{thm:juntaness-tolerant}
    For all \(0 \le \eps_1 < \eps_2 < 1\), $\delta \in (0,1/2)$, \(k \in \Zbb_{\ge 1}\), there is an $(\eps_1, \eps_2, \delta)$-tester for being a $k$-junta using \[
        \bO{\frac{k}{(\eps_2-\eps_1)^2}
        \log^2\left(\frac{k}{\delta(\eps_2-\eps_1)}\right)
        \log^5\left(\frac{2}{\eps_2-\eps_1}\right)}
    \]
    total Hamiltonian evolution time and
    \[
        \bO{\frac{k}{\eps_2(\eps_2-\eps_1)^3}
        \log^2\left(\frac{k}{\delta(\eps_2-\eps_1)}\right)
        \log^9\left(\frac{2}{\eps_2-\eps_1}\right)}
    \]
    queries to the time evolution operator.
\end{restatable}

\begin{proof}
    For each \(J\subseteq[n]\), define
    \begin{align*}
        G_J
        =
        \left\{
            P_1\otimes\cdots\otimes P_n\in\Pbb_n:
            P_i=I \text{ for every } i\notin J, P_i\in\{I, X, Y, Z\} \text{ for every } i\in J
        \right\}.
    \end{align*}
    Take \(S_\text{junta} = \{G_J: J\subseteq[n]\}\) and \(\ell(G_J) =
    \abs{J}\). This satisfies the requirements of \cref{dfn:gendim}:
    $G_\emptyset$ is the unique zero-dimensional group, and for all $G_J$ and
    $P \notin G_J$, the minimal $G_{J'}$ containing $G_J$ and $P$ is
    given by setting $J'$ to the union of $J$ and $\set{i \in \brac{n} : P_i
    \not= I}$.  $H$ is a $k$-junta iff $\supp(H)$ is contained in $G_J$ for some
    $J$ with $\ell(G_J) = \abs{J} \le k$, so we may apply
    \cref{thm:group-property-tester}.
\end{proof}

We now consider the Pauli dimensionality testing problem. 

\begin{restatable}{theorem}{paulidimensionality-tolerant}
  \label{thm:paulidimensionality-tolerant}
    For all \(0 \le \eps_1 < \eps_2 < 1\), $\delta \in (0,1/2)$, there is an $(\eps_1, \eps_2, \delta)$-tester for
    having Pauli dimensionality $k \in \Zbb_{\ge 1}$ that uses \[
        \bO{\frac{k}{(\eps_2-\eps_1)^2}
        \log^2\left(\frac{k}{\delta(\eps_2-\eps_1)}\right)
        \log^5\left(\frac{2}{\eps_2-\eps_1}\right)}
    \]
    total Hamiltonian evolution time and
    \[
        \bO{\frac{k}{\eps_2(\eps_2-\eps_1)^3}
        \log^2\left(\frac{k}{\delta(\eps_2-\eps_1)}\right)
        \log^9\left(\frac{2}{\eps_2-\eps_1}\right)}
    \]
    queries to the time evolution operator.
\end{restatable}

\begin{proof}
    Define $\Sc$ to be the set of all subgroups of $\Pbb_n$, and $\ell(G)$ as the
    size of the minimum generating set for $G$. The unique $\ell$-minimal element
    is then the trivial group. $\Pb_n$ is isomorphic to $\Fbb_4^{n} \cong
    \Zbb_2^{2n}$ (as quotienting out phases makes the group Abelian), and so for
    every subgroup $G$, adding an element $P$ from outside of $G$ is guaranteed to
    increase $\ell(G)$, with the unique minimal $G'$ containing $P$ and all of
    $G$ being the group generated by $P$ and $G$. So $(S,\ell)$ satisfy the
    requirements of \cref{dfn:gendim}. By definition, $H$ has Pauli dimensionality at most k iff it has $(S,\ell)$-dimensionality at most $k$, so we may apply \cref{thm:group-property-tester}.
\end{proof}

\subsection{Locality Tester}

We first define locality.

\begin{definition}[\(k\)-locality]
    A Hamiltonian \(H\) is said to be \(k\)-local if every Pauli in its support acts non-trivially on at most \(k\) qubits.
\end{definition}

The principal result of this subsection is an algorithm for the tolerant locality testing problem.

\begin{restatable}{theorem}{locality-tolerant}\label{thm:locality-tester-tolerant}
    For all \(0 \le \eps_1 < \eps_2 < 1\), \(\delta \in (0,1/2)\), there is an \((\eps_1, \eps_2, \delta)\)-tester for \(k\)-locality, where \(k \in \Zbb_{\ge 0}\), that uses
    \[
        \bO{\frac{1}{(\eps_2-\eps_1)^2}
        \log{\left(\frac{1}{\delta}\right)}
        \log\left(\frac{1}{\delta(\eps_2-\eps_1)}\right)
        \log^5\left(\frac{2}{\eps_2-\eps_1}\right)}
    \]
    total Hamiltonian evolution time and
    \[
        \bO{\frac{1}{\eps_2(\eps_2-\eps_1)^3}
        \log{\left(\frac{1}{\delta}\right)}
        \log\left(\frac{1}{\delta(\eps_2-\eps_1)}\right)
        \log^9\left(\frac{2}{\eps_2-\eps_1}\right)}
    \]
    queries to the time evolution operator.
\end{restatable}

Let \(S_k \subseteq \Pbb_n\) be the set of Paulis that act non-trivially on at most \(k\) qubits. We now describe the algorithm:
\begin{enumerate}
    \item Using~\cref{lem:norm-estimate}, with probability at least \(1-\frac{\delta}{3}\), we either learn that \(\norm{H}_2^2 < \eps_2^2\), or we obtain a \(\left(1 \pm \frac{\eps_2^2-\eps_1^2}{100\eps_2^2}\right)\)-approximation \(h\) of \(\norm{H}_2^2\). If \(\norm{H}_2^2 < \eps_2^2\), accept. Otherwise, continue.

    \item Set
    \[
        N=\ceil*{
            \frac{200000\eps_2^2h}{(\eps_2^2-\eps_1^2)^2}
            \log\left(\frac{6}{\delta}\right)}.
    \]
    Obtain \(N\) independent samples using~\cref{thm:spectrum-sampling-H}, with accuracy \(\gamma=\frac{\eps_2^2-\eps_1^2}{100}\) and failure probability \(\frac{\delta}{3N}\) for each sample.

    \item Let \(\widehat p\) be the fraction of samples outside \(S_k\), and set \(\widehat T(S_k)=h\widehat p\). The algorithm accepts iff
    \[
        \widehat T(S_k) \le \frac{\eps_1^2+\eps_2^2}{2}.
    \]
\end{enumerate}

\begin{proof}[Proof of~\cref{thm:locality-tester-tolerant}]
    By Parseval's theorem, the squared distance of \(H\) from being \(k\)-local is
    \[
        T(S_k)=\sum_{P \in \Pbb_n\setminus S_k}\abs{\alpha_P}^2.
    \]
    If~\cref{lem:norm-estimate} certifies that \(\norm{H}_2^2 < \eps_2^2\), we may accept, since the zero Hamiltonian is \(k\)-local. Otherwise, let \(h\) be the returned approximation, with \(h \ge \frac{\eps_2^2}{2}\). Condition on this stage succeeding and on no sampling failures. The sampling failure probability bounds the chance of exceeding the stated costs of sampling. A union bound bounds the probability of either failure by \(\frac{2\delta}{3}\).

    Let \(\widetilde p\) be the probability that a sample lies outside \(S_k\). By the total variation distance guarantee of~\cref{thm:spectrum-sampling-H},
    \[
        \abs*{\widetilde p-\frac{T(S_k)}{\norm{H}_2^2}} \le \gamma.
    \]
    Consequently,
    \[
        \abs{h\widetilde p-T(S_k)}
        \le \frac{\eps_2^2-\eps_1^2}{100\eps_2^2}T(S_k)+h\gamma.
    \]
    Since \(h \le \frac{101}{100}\norm{H}_2^2 \le \frac{101}{100}\), this also gives
    \[
        h\widetilde p \le 2\max\set*{T(S_k),\eps_2^2}.
    \]
    By~\cref{lem:bernstein-prelim} and the choice of \(N\), with probability at least \(1-\frac{\delta}{3}\),
    \[
        h\abs{\widehat p-\widetilde p}
        \le \frac{\eps_2^2-\eps_1^2}{100\eps_2^2}
            \max\set*{T(S_k),\eps_2^2}.
    \]
    Combining these inequalities gives
    \begin{align*}
        \abs{\widehat T(S_k)-T(S_k)}
        &\le h\abs{\widehat p-\widetilde p}
            +\abs{h\widetilde p-T(S_k)}\\
        &<\frac{\eps_2^2-\eps_1^2}{10\eps_2^2}
            \max\set*{T(S_k),\eps_2^2}.
    \end{align*}
    Condition on this event as well. All stages succeed with probability at least \(1-\delta\).

    For completeness, suppose \(H\) is \(\eps_1\)-close to being \(k\)-local. Then \(T(S_k) \le \eps_1^2\), so
    \[
        \widehat T(S_k)
        \le \eps_1^2+\frac{\eps_2^2-\eps_1^2}{10}
        <\frac{\eps_1^2+\eps_2^2}{2}.
    \]
    Hence the tester accepts.

    For soundness, suppose \(H\) is \(\eps_2\)-far from being \(k\)-local. Then \(T(S_k) \ge \eps_2^2\), so
    \begin{align*}
        \widehat T(S_k)
        &\ge \left(1-\frac{\eps_2^2-\eps_1^2}{10\eps_2^2}\right)T(S_k)\\
        &\ge \eps_2^2-\frac{\eps_2^2-\eps_1^2}{10}
        >\frac{\eps_1^2+\eps_2^2}{2}.
    \end{align*}
    Hence the tester rejects.

    Since \(h=\Theta(\norm{H}_2^2)\) and \(h \ge \frac{\eps_2^2}{2}\),~\cref{thm:spectrum-sampling-H} gives total evolution time
    \begin{align*}
        &\bO{\frac{N}{\norm{H}_2^2}
            \log\left(\frac{3N}{\delta}\right)
            \log^5\left(\frac{2}{\gamma}\right)}\\
        &=\bO{\frac{\eps_2^2}{(\eps_2^2-\eps_1^2)^2}
            \log{\left(\frac{1}{\delta}\right)}
            \log\left(\frac{1}{\delta(\eps_2-\eps_1)}\right)
            \log^5\left(\frac{2}{\eps_2-\eps_1}\right)}
    \end{align*}
    and query complexity
    \begin{align*}
        &\bO{\frac{N}{\norm{H}_2^2\gamma}
            \log\left(\frac{3N}{\delta}\right)
            \log^9\left(\frac{2}{\gamma}\right)}\\
        &=\bO{\frac{\eps_2^2}{(\eps_2^2-\eps_1^2)^3}
            \log{\left(\frac{1}{\delta}\right)}
            \log\left(\frac{1}{\delta(\eps_2-\eps_1)}\right)
            \log^9\left(\frac{2}{\eps_2-\eps_1}\right)}.
    \end{align*}
    These costs subsume those in~\cref{lem:norm-estimate}. Using \(\eps_2(\eps_2-\eps_1) \le \eps_2^2-\eps_1^2 \le 2\eps_2(\eps_2-\eps_1)\) gives the claimed bounds.
\end{proof}

\subsection{Hamiltonian Certification}

In this subsection, we give an algorithm for Hamiltonian certification. Throughout this subsection, \(H_0\) is the target Hamiltonian, where \(\norm{H_0}_\infty\le1\) and \(H_0\) is traceless.

\begin{definition}[Hamiltonian certification]
    For a target Hamiltonian \(H_0\), the certification property is \(\Pc_{H_0}\coloneqq\{H_0\}\).
\end{definition}

The principal result of this subsection is an algorithm for the tolerant Hamiltonian certification problem.

\begin{restatable}{theorem}{HamiltonianCertificationTester}
\label{thm:hamiltonian-certification}
    For all \(0<\eps_1<\eps_2<1\), \(\delta\in(0,1/2)\), there is an \((\eps_1,\eps_2,\delta)\)-tester for \(\Pc_{H_0}\) that uses
    \[
        \bO{\frac{1}{(\eps_2-\eps_1)^2}
        \log{\left(\frac1\delta\right)}
        \log^5\left(\frac{2}{\eps_2-\eps_1}\right)}
    \]
    total Hamiltonian evolution time and
    \[
        \bO{\frac{1}{(\eps_2-\eps_1)^3}
        \log{\left(\frac1\delta\right)}
        \log^9\left(\frac{2}{\eps_2-\eps_1}\right)}
    \]
    queries to the time evolution operator.
\end{restatable}

We now describe the algorithm:
\begin{enumerate}
    \item Set \(\gamma=\frac{\eps_2-\eps_1}{4}\). Obtain an \((\alpha,a,\gamma)\)-block encoding of \(H-H_0\), with \(2 \le \alpha = \bO{\log^2{\left(\frac{2}{\gamma}\right)}}\) and \(a = \bO{n + \log{\left(\frac{2}{\gamma}\right)}}\).

    \item For \(N=\ceil*{\frac{128\alpha^2}{(\eps_2-\eps_1)^2}\log\left(\frac2\delta\right)}\) iterations, attempt to prepare the pseudo-Choi state \(\ket{\Psi(\widetilde{A})}\) using the block encoding obtained, where \(\widetilde{A}\) is an operator such that \(\norm{\widetilde A - (H-H_0)}_\infty \le \gamma\). Let \(\widehat p\) be the fraction of successful attempts. The algorithm accepts iff
    \[
        \widehat p\le
        \frac{(\eps_1+\gamma)^2+(\eps_2-\gamma)^2}{2\alpha^2}.
    \]
\end{enumerate}

We first state the singular value amplification result that we use to obtain a constant normalization factor in the block encoding of \(H_0\).

\begin{lemma}\cite[Theorem 30]{GSLW19}\label{lem:singular-value-amplification}
    Let \(U\) be an \((\alpha,a,0)\)-block encoding of an \(n\)-qubit operator \(A\), where \(\alpha>0\). Let \(\lambda>1\) and \(\eta,\eps\in(0,1/2)\), and suppose that
    \[
        \norm{A}_\infty\le\frac{\alpha(1-\eta)}{\lambda}.
    \]
    We can construct an \(\left(\frac{\alpha}{\lambda},a+1,\frac{\alpha\eps}{\lambda}\right)\)-block encoding of \(A\), using
    \[
        \bO{\frac{\lambda}{\eta}\log\left(\frac{\lambda}{\eps}\right)}
    \]
    applications of \(U\) and \(U^\dagger\).
\end{lemma}

We now show that we can block encode the known Hamiltonian \(H_0\) with a constant normalization factor.

\begin{lemma}\label{lem:block-encoding-known-H}
    For any Hamiltonian \(H_0 = \sum_{P \in \Pbb_n} \alpha'_P P\) such that \(\norm{H_0}_\infty\le1\), and \(\gamma\in(0,1)\), we can construct a \((2,\bO{\log(M+2)},\gamma)\)-block encoding of \(H_0\), where \(M\) is the number of non-zero Pauli coefficients \(\alpha'_P\).
\end{lemma}

\begin{proof}
    Write \(H_0=\sum_{j=1}^M\beta_jP_j\) and \(\Lambda=\sum_{j=1}^M\abs{\beta_j}\). If \(H_0=0\), the Pauli \(X\) on one ancilla qubit gives the required block encoding. Otherwise, using~\cref{lem:linear-combination-be}, obtain a \((\Lambda,a,0)\)-block encoding \(B\) of \(H_0\), where \(a=\ceil*{\log_2 M}\).

    If \(\Lambda\le2\), append one ancilla qubit and apply a single-qubit rotation \(R\) satisfying
    \[
        R\ket0=\frac{\Lambda}{2}\ket0
        +\sqrt{1-\frac{\Lambda^2}{4}}\ket1.
    \]
    The zero block of \(R\otimes B\) is
    \[
        (\bra{0^{a+1}}\otimes I)(R\otimes B)
        (\ket{0^{a+1}}\otimes I)
        =\frac{\Lambda}{2}\cdot\frac{H_0}{\Lambda}
        =\frac{H_0}{2}.
    \]
    Therefore \(R\otimes B\) is a \((2,a+1,0)\)-block encoding of \(H_0\).

    If \(\Lambda>2\), apply~\cref{lem:singular-value-amplification} with \(\lambda=\frac\Lambda2\), \(\eta=\frac14\), and \(\eps=\frac\gamma2\). We can apply~\cref{lem:singular-value-amplification} since
    \[
        \norm*{\frac{H_0}{\Lambda}}_\infty
        \le\frac1\Lambda
        \le\frac{1-\frac14}{\frac\Lambda2}.
    \]
    This gives a \((2,a+1,\gamma)\)-block encoding of \(H_0\), using \(\bO{\Lambda\log\left(\frac{2\Lambda}{\gamma}\right)}\) applications of \(B\) and \(B^\dagger\).
\end{proof}

We now combine this construction with the block encoding procedure of~\cref{lem:block-encoding-H} to block encode \(H-H_0\).

\begin{lemma}\label{lem:block-encoding-difference}
    For any Hamiltonians \(H\) and \(H_0\) such that \(\norm{H}_\infty,\norm{H_0}_\infty\le1\), and \(\gamma\in(0,1)\), we can construct an \((\alpha,a,\gamma)\)-block encoding of \(H-H_0\), where
    \(
        2\le\alpha=\bO{\log^2\left(\frac2\gamma\right)},
    \) and \(
        a=\bO{n+\log\left(\frac2\gamma\right)}.
    \)
    This requires \(\bO{\frac{1}{\gamma}\log^5\left(\frac{2}{\gamma}\right)}\) queries to the time evolution operator of \(H\) and \(\bO{\log\left(\frac{2}{\gamma}\right)}\) total evolution time.
\end{lemma}

\begin{proof}
    Using~\cref{lem:block-encoding-H,lem:block-encoding-known-H} with accuracy \(\frac\gamma4\), obtain block encodings of \(H\) and \(H_0\). Let \(\Lambda\ge2\) be the larger normalization factor.

    To increase the smaller normalization factor \(\beta\) to \(\Lambda\), append one ancilla qubit and tensor the corresponding block encoding with a single-qubit rotation \(R\) satisfying
    \[
        R\ket0=\frac{\beta}{\Lambda}\ket0
        +\sqrt{1-\frac{\beta^2}{\Lambda^2}}\ket1.
    \]
    If \(C\) is the original zero block and \(A\) is the operator it encodes, the new zero block is \(\frac{\beta}{\Lambda}C\), and
    \[
        \norm*{A-\Lambda\frac{\beta}{\Lambda}C}_\infty
        =\norm{A-\beta C}_\infty
        \le\frac\gamma4.
    \]
    Therefore the normalization increases to \(\Lambda\) without increasing the error or making any additional evolution queries. Padding the ancilla registers if necessary, both block encodings are now \((\Lambda,a',\frac\gamma4)\)-block encodings.

    Let \(B\) and \(B'\) denote the block encodings of \(H\) and \(H_0\), with zero blocks \(\frac{\widetilde H}{\Lambda}\) and \(\frac{\widetilde H_0}{\Lambda}\), respectively. Use disjoint ancilla registers for these block encodings.

    The construction in~\cref{lem:block-encoding-H} can be modified to obtain \(\ket0\bra0\otimes B+\ket1\bra1\otimes K\), by replacing each control string \(b\) by \(0b\) in the applications of~\cref{lem:controlization}, and controlling the known gates in the construction. Here, \(K\) is an operator that has a zero block \(K'\) satisfying \(\norm{K'-I}_\infty\le\frac\gamma4\). This preserves the approximation guarantees of \(B\) as well as the asymptotic costs. The block encoding of \(H_0\) can be controlized to obtain \(\ket0\bra0\otimes I+\ket1\bra1\otimes B'\), by controlling its known gates.

    To apply~\cref{lem:linear-combination-be} with coefficients \((1,-1)\), choose unitaries
    \((\texttt{PREP}_L,\texttt{PREP}_R)\) such that
    \begin{align*}
        \texttt{PREP}_R\ket0
        &=\ket{+}\\
        \texttt{PREP}_L\ket0
        &=\ket{-}.
    \end{align*}
    Let
    \begin{align*}
        \texttt{SEL}
        &=(\ket0\bra0\otimes I+\ket1\bra1\otimes B')
          (\ket0\bra0\otimes B+\ket1\bra1\otimes K)\\
        &=\ket0\bra0\otimes B+\ket1\bra1\otimes B'K.
    \end{align*}
    Apply \(\texttt{PREP}_R\otimes I\), then \(\texttt{SEL}\), and then \(\texttt{PREP}_L^\dagger\otimes I\). Since the ancilla registers are disjoint, the resulting zero block is
    \[
        \frac12\left(
            \frac{\widetilde H}{\Lambda}
            -\frac{\widetilde H_0K'}{\Lambda}
        \right).
    \]
    Therefore the normalization is \(\alpha=2\Lambda\), and the
    block-encoding error is at most
    \[
        \norm{H-\widetilde H}_\infty
        +\norm{H_0-\widetilde H_0}_\infty
        +\norm{\widetilde H_0}_\infty\norm{I-K'}_\infty
        \le\frac\gamma4+\frac\gamma4
        +\left(1+\frac\gamma4\right)\frac\gamma4
        \le\gamma.
    \]

    The resource bounds follow from~\cref{lem:block-encoding-H,lem:block-encoding-known-H}.
\end{proof}

We are now ready to prove our theorem.

\begin{proof}[Proof of~\cref{thm:hamiltonian-certification}]
    Set \(\gamma=\frac{\eps_2-\eps_1}{4}\), and obtain a block encoding \(B\) of \(H-H_0\) using \cref{lem:block-encoding-difference}. Write its zero block as \(\frac{\widetilde A}{\alpha}\).

    For each preparation attempt of \(\ket{\Psi(\widetilde{A})}\), consider two \(n\)-qubit work registers \(\Ac,\Bc\), and an \(a\)-qubit ancilla register \(\Cc\). Prepare \(\ket\Phi\) in \(\Ac\Bc\) and \(\ket{0^a}\) in \(\Cc\). Apply \(B\) to \(\Cc\Bc\), and measure \(\Cc\). Declare success iff the outcome is \(0^a\). The unnormalized state on \(\Ac\Bc\) corresponding to success is
    \[
        \frac{(I\otimes\widetilde A)\ket\Phi}{\alpha}.
    \]
    Hence, the success indicator is a Bernoulli random variable with mean
    \[
        p=\frac{\norm{\widetilde A}_2^2}{\alpha^2}.
    \]
    Therefore each attempt gives an independent sample from this Bernoulli random variable.

    Since \(B\) is an \((\alpha,a,\gamma)\)-block encoding of \(H-H_0\), the reverse triangle inequality gives
    \[
        \abs{\norm{\widetilde A}_2-\norm{H-H_0}_2}
        \le\norm{\widetilde A-(H-H_0)}_2
        \le\norm{\widetilde A-(H-H_0)}_\infty
        \le\gamma.
    \]
    Consequently,
    \[
        \abs*{\sqrt p-\frac{\norm{H-H_0}_2}{\alpha}}
        \le\frac{\gamma}{\alpha}.
    \]
    Define
    \begin{align*}
        \eps_1'=\frac{\eps_1+\gamma}{\alpha},
        \\
        \eps_2'=\frac{\eps_2-\gamma}{\alpha}.
    \end{align*}
    Since \(\alpha\ge2\), we have \(0<\eps_1'<\eps_2'<1\), and
    \[
        \eps_2'-\eps_1'
        =\frac{\eps_2-\eps_1-2\gamma}{\alpha}
        =\frac{\eps_2-\eps_1}{2\alpha}.
    \]
    The certification problem now reduces to distinguishing between \(p\le(\eps_1')^2\) and \(p\ge(\eps_2')^2\). We verify that the stated sample count and threshold suffice for this distinguishing task. The probability of rejection is non-decreasing in \(p\), so it suffices to analyze \(p=(\eps_1')^2\) and \(p=(\eps_2')^2\). Set
    \[
        \eta=\frac{(\eps_2')^2-(\eps_1')^2}{3}.
    \]
    For either of these means, \(p\le(\eps_2')^2\), and~\cref{lem:bernstein-prelim} gives
    \begin{align*}
        \prob{\abs{\widehat p-p}>\eta}
        &\le2\exp\left(
            -\frac{N\eta^2}{2(\eps_2')^2+2\eta/3}
        \right)\\
        &\le2\exp\left(
            -\frac{N((\eps_2')^2-(\eps_1')^2)^2}
            {20(\eps_2')^2}
        \right)\\
        &\le2\exp\left(
            -\frac{N(\eps_2'-\eps_1')^2}{20}
        \right)
        \le\delta.
    \end{align*}
    The last inequality follows from the choice of \(N\). On the complementary event, \(\widehat p\) lies on the correct side of the threshold
    \[
        \frac{(\eps_1')^2+(\eps_2')^2}{2}
        =\frac{(\eps_1+\gamma)^2+(\eps_2-\gamma)^2}{2\alpha^2}.
    \]

    For completeness, suppose \(H\) is \(\eps_1\)-close to \(\Pc_{H_0}\). Then \(\norm{H-H_0}_2\le\eps_1\), so
    \[
        \sqrt p
        \le\frac{\norm{H-H_0}_2+\gamma}{\alpha}
        \le\eps_1'.
    \]
    Hence the tester accepts with probability at least \(1-\delta\).

    For soundness, suppose \(H\) is \(\eps_2\)-far from \(\Pc_{H_0}\). Then \(\norm{H-H_0}_2\ge\eps_2\), so
    \[
        \sqrt p
        \ge\frac{\norm{H-H_0}_2-\gamma}{\alpha}
        \ge\eps_2'.
    \]
    Hence the tester rejects with probability at least \(1-\delta\).

    Finally, the number of preparation attempts is
    \[
        N=\bO{\frac{\alpha^2}{(\eps_2-\eps_1)^2}
        \log\left(\frac1\delta\right)}.
    \]
    Each attempt requires \(\bO{\frac{1}{\gamma}\log^5\left(\frac{2}{\gamma}\right)}\) queries to the time evolution operator of \(H\) and \(\bO{\log\left(\frac{2}{\gamma}\right)}\) total evolution time. Since \(\alpha=\bO{\log^2\left(\frac2\gamma\right)}\) and \(\gamma=\frac{\eps_2-\eps_1}{4}\), the tester uses
    \[
        \bO{\frac1{(\eps_2-\eps_1)^2}
        \log\left(\frac1\delta\right)
        \log^5\left(\frac{2}{\eps_2-\eps_1}\right)}
    \] total Hamiltonian evolution time and
    \[
        \bO{\frac1{(\eps_2-\eps_1)^3}
        \log\left(\frac1\delta\right)
        \log^9\left(\frac{2}{\eps_2-\eps_1}\right)}
    \] queries to the time evolution operator of \(H\).
\end{proof}

\section{Lower Bounds}
\label{sec:lower-bounds}

In this section, we show that tolerantly testing locality, sparsity, degree-boundedness, juntaness, and Pauli dimensionality requires
\[
    \bOm{\min\set*{\frac{1}{\eps_2(\eps_2 - \eps_1)},\frac{2^{\frac{n}{2}}}{\sqrt{\eps_2(\eps_2 - \eps_1)}}}}
\]
total forward evolution time.  We use the random Hamiltonian ensemble from~\cite{GLZ26} to establish these lower bounds. 

For \(0<\alpha<\frac{1}{2}\), let \(\nu_\alpha\) be the distribution from~\cite[Definition 3.1]{GLZ26}, and draw
\begin{align*}
    H=\sum_{x\in\bool^n}h_x\ketbra{x}{x},
    \qquad h_x\stackrel{\mathrm{iid}}{\sim}\nu_\alpha.
\end{align*}
By~\cite[Lemma 3.2]{GLZ26},
\begin{align*}
    \abs{h_x}\le1,
    \qquad \E{h_x}=0,
    \qquad \E{h_x^2}=\alpha^2.
\end{align*}
In particular, \(\norm{H}_\infty\le1\). By~\cite[Theorem 3.10]{GLZ26}, any algorithm whose acceptance probabilities on the zero Hamiltonian and on this random \(H\) differ by a positive constant requires \(\bOm{\frac{1}{\alpha^2}}\) total forward evolution time.

For \(S\subseteq[n]\), let \(Z_S\) be the Pauli string with \(Z\) on \(S\) and \(I\) elsewhere. Write
\begin{align*}
    H=\sum_{S\subseteq[n]}a_SZ_S,
    \qquad
    \text{where}~a_S=\frac{1}{2^n}\sum_{x\in\bool^n}
        h_x(-1)^{\sum_{j\in S}x_j}.
\end{align*}

We first show that the coefficients involving the first qubit retain substantial squared weight after any small set of coefficients is removed.

\begin{lemma}\label{lem:random-farness}
    Assume that \(\alpha^2\ge\frac{256}{2^n}\). With probability at least \(\frac{2}{3}\), for every \(\Ac\subseteq2^{[n]}\) with \(\abs{\Ac}\le\frac{2^n}{512}\),
    \begin{align*}
        \sum_{\substack{S\subseteq[n]\\1\in S,\ S\notin\Ac}}a_S^2
        >\frac{\alpha^2}{4}.
    \end{align*}
\end{lemma}

\begin{proof}
    Write \(Q_1=\sum_{S\ni1}a_S^2\). The Fourier coefficients of the function \(y\mapsto\frac{h_{0y}-h_{1y}}{2}\), for \(y\in\bool^{n-1}\), are \(a_{T\cup\{1\}}\), for \(T\subseteq\{2,\dots,n\}\). Thus, Parseval's identity gives
    \begin{align*}
        Q_1=\frac{1}{2^{n+1}}
            \sum_{y\in\bool^{n-1}}(h_{0y}-h_{1y})^2.
    \end{align*}
    The summands use disjoint pairs of independent variables \(h_x\), and
    \begin{align*}
        \E{(h_{0y}-h_{1y})^2}&=2\alpha^2,\\
        \E{(h_{0y}-h_{1y})^4}&=2\E{h_x^4}+6\alpha^4.
    \end{align*}
    Consequently,
    \begin{align*}
        \E{Q_1}&=\frac{\alpha^2}{2},\\
        \Var{Q_1}
        &=\frac{\E{h_x^4}+\alpha^4}{2^{n+2}}
        \le\frac{5\alpha^2}{2^{n+4}},
    \end{align*}
    where we used \(\E{h_x^4}\le\E{h_x^2}=\alpha^2\) and \(\alpha^2<\frac{1}{4}\). By Chebyshev's inequality,
    \begin{align*}
        \prob{Q_1\le\frac{3\alpha^2}{8}}
        \le\frac{\Var{Q_1}}{\left(\frac{\alpha^2}{8}\right)^2}
        \le\frac{20}{2^n\alpha^2}
        \le\frac{5}{64}.
    \end{align*}

    We next bound the weight that can be removed. Expanding the fourth power in the definition of \(a_S\), independence and \(\E{h_x}=0\) leave only terms with all four indices equal or with two equal pairs. The signs multiply to one in both cases, so
    \begin{align*}
        \E{a_S^4}
        &=\frac{2^n\E{h_x^4}+3\cdot2^n(2^n-1)\alpha^4}{2^{4n}}\\
        &\le\frac{\alpha^2}{2^{3n}}+\frac{3\alpha^4}{2^{2n}}.
    \end{align*}
    There are \(2^{n-1}\) sets \(S\) containing the first qubit. Hence,
    \begin{align*}
        \E{\sum_{S\ni1}a_S^4}
        \le\frac{\alpha^4}{2^{n+1}}
            \paren*{3+\frac{1}{2^n\alpha^2}}
        \le\frac{2\alpha^4}{2^n}.
    \end{align*}
    For every \(\Ac\) with \(\abs{\Ac}\le\frac{2^n}{512}\), Cauchy--Schwarz gives
    \begin{align*}
        \left(\sum_{\substack{S\in\Ac\\1\in S}}a_S^2\right)^2
        \le\abs{\Ac}\sum_{S\ni1}a_S^4
        \le\frac{2^n}{512}\sum_{S\ni1}a_S^4.
    \end{align*}
    Thus, by Markov's inequality,
    \begin{align*}
        \prob{\max_{\substack{\Ac\subseteq2^{[n]}\\\abs{\Ac}\le\frac{2^n}{512}}}
            \sum_{\substack{S\in\Ac\\1\in S}}a_S^2
            \ge\frac{\alpha^2}{8}}
        &\le\frac{64}{\alpha^4}\frac{2^n}{512}
            \E{\sum_{S\ni1}a_S^4}\\
        &\le\frac{1}{4}.
    \end{align*}
    Outside these two failure events, every such \(\Ac\) satisfies
    \begin{align*}
        \sum_{\substack{S\subseteq[n]\\1\in S,\ S\notin\Ac}}a_S^2
        =Q_1-\sum_{\substack{S\in\Ac\\1\in S}}a_S^2
        >\frac{3\alpha^2}{8}-\frac{\alpha^2}{8}
        =\frac{\alpha^2}{4}.
    \end{align*}
    The combined failure probability is at most \(\frac{5}{64}+\frac{1}{4}=\frac{21}{64}<\frac{1}{3}\), proving the claim.
\end{proof}

\begin{theorem}\label{thm:tolerant-property-testing-lower-bounds}
    For all \(n \ge 12\) and \(0 < \eps_1 < \eps_2 < \frac{1}{32}\), any \((\eps_1, \eps_2, 1/3)\)-tester for any of the following properties requires
    \[
        \bOm{\min\set*{
            \frac{1}{\eps_2(\eps_2-\eps_1)},
            \frac{2^{\frac{n}{2}}}{\sqrt{\eps_2(\eps_2-\eps_1)}}}}
    \]
    total forward Hamiltonian evolution time:
    \begin{enumerate}[label=(\roman*)]
        \item \(k\)-locality, for \(k \in \Zbb_{\ge 0}\) with \(k \le \frac{n}{16}\).
        \item \(s\)-sparsity, for \(s \in \Zbb_{\ge 0}\) with \(s \le \frac{2^n}{1024}\).
        \item \(\Delta\)-degree-boundedness, for \(\Delta \in \Zbb_{\ge 0}\) with \(\Delta \le \frac{2^n}{1024}\).
        \item \(k\)-juntaness, for \(k \in \Zbb_{\ge 0}\) with \(k \le n-10\).
        \item \(k\)-Pauli dimensionality, for \(k \in \Zbb_{\ge 0}\) with \(k \le n-10\).
        \item  Equality to a known Hamiltonian \(H_0\) (Hamiltonian certification).
    \end{enumerate}
\end{theorem}

\begin{proof}[Proof of~\cref{thm:tolerant-property-testing-lower-bounds}]
    Let \(\Pc\) be any of the properties listed in the theorem, and let \(\Pc'\) be the same property, with the same parameter, on \(n-1\) qubits. For any \(K \in \Pc\), write
    \[
        K=K_I\otimes I+K_X\otimes X+K_Y\otimes Y+K_Z\otimes Z.
    \]
    We have \(K_Z\otimes Z \in \Pc\) and \(K_I,K_Z \in \Pc'\), since deleting Pauli terms and removing the last tensor factor cannot increase any of the five parameters.

    For any \(L \in \Pc'\), let
    \[
        \Ac(L)=\set{S \subseteq [n-1]:1 \in S,\ \wh{L}(Z_S) \neq 0}.
    \]
    Then,
    \[
        \abs{\Ac(L)} \le
        \begin{cases}
            \displaystyle\sum_{j=1}^k\binom{n-2}{j-1}, & \text{for locality},\\
            s, & \text{for sparsity},\\
            \Delta, & \text{for degree-boundedness},\\
            2^k, & \text{for juntaness or Pauli dimensionality}
        \end{cases}
        \le \frac{2^{n-1}}{512}.
    \]
    For juntaness, there are at most \(2^k\) strings containing only \(I\)'s and \(Z\)'s on \(k\) qubits. For Pauli dimensionality, a group generated by at most \(k\) Paulis contains at most \(2^k\) elements.

    Let \(H=\sum_{S \subseteq [n-1]}a_SZ_S\) be drawn from the ensemble above on \(n-1\) qubits, with \(\alpha^2 \ge \frac{256}{2^{n-1}}\). By~\cref{lem:random-farness}, with probability at least \(\frac{2}{3}\), simultaneously for every \(L \in \Pc'\),
    \begin{equation}\label{eq:random-property-farness}
        \sum_{\substack{S \subseteq [n-1]\\1 \in S,\ \wh{L}(Z_S)=0}}a_S^2
        >\frac{\alpha^2}{4}.
    \end{equation}
    Applying this with \(\alpha=\frac{3}{8}\), fix a diagonal Hamiltonian \(D\) such that \(\norm{D}_\infty \le 1\) and \(\norm{D-L}_2>\frac{3}{16}\) for every \(L \in \Pc'\). This choice of \(\alpha\) is valid since \(n \ge 12\). Set
    \[
        d=\min_{K \in \Pc}\norm{D\otimes Z-K}_2,
        \qquad A=\frac{D}{d}.
    \]
    Since \(K_Z \in \Pc'\), Pauli orthogonality gives
    \[
        \norm{D\otimes Z-K}_2 \ge \norm{D-K_Z}_2>\frac{3}{16}.
    \]
    Therefore, \(d \ge \frac{3}{16}\), \(\norm{A}_\infty \le 16\), and, since \(\Pc\) is closed under rescaling,
    \[
        \min_{K \in \Pc}\norm{A\otimes Z-K}_2=1.
    \]

    Set
    \begin{align*}
        \eps&=\sqrt{\eps_2^2-\eps_1^2},
        &\alpha&=\max\set*{8\eps,\frac{16}{2^{\frac{n-1}{2}}}},\\
        \widetilde H&=\frac{2\eps}{\alpha}(H-a_\emptyset I),
        &B&=\eps_1 A\otimes Z+\widetilde H\otimes I,
    \end{align*}
    where \(H\) is drawn from the ensemble with parameter \(\alpha\), and \(a_\emptyset=2^{-(n-1)}\Tr(H)\). Our assumptions give \(0 < \alpha < \frac{1}{2}\) and \(\alpha^2 \ge \frac{256}{2^{n-1}}\). Also, \(B\) is traceless, and
    \[
        \norm{B}_\infty
        \le 16\eps_1+\frac{2\eps}{\alpha}
            \paren*{\norm{H}_\infty+\abs{a_\emptyset}}
        \le 16\eps_1+\frac{4\eps}{\alpha}<1.
    \]

    When \(H=0\), the distance of \(B=\eps_1 A\otimes Z\) from \(\Pc\) is exactly \(\eps_1\). When~\cref{eq:random-property-farness} holds, every \(L \in \Pc'\) satisfies
    \[
        \norm{\widetilde H-L}_2^2
        \ge \frac{4\eps^2}{\alpha^2}
            \sum_{\substack{S \subseteq [n-1]\\1 \in S,\ \wh{L}(Z_S)=0}}a_S^2
        >\eps^2.
    \]
    Consequently, for every \(K \in \Pc\), Pauli orthogonality gives
    \begin{align*}
        \norm{B-K}_2^2
        &\ge \norm{\eps_1 A-K_Z}_2^2
            +\norm{\widetilde H-K_I}_2^2\\
        &\ge \eps_1^2+\eps^2=\eps_2^2.
    \end{align*}
    Thus, a tolerant tester rejects with probability at most \(\frac{1}{3}\) when \(H=0\), and with probability at least \(\frac{2}{3}\cdot\frac{2}{3}=\frac{4}{9}\) when averaged over random \(H\).

    Since \(A\) and \(H\) are diagonal,
    \[
        e^{-itB}
        =e^{-it\eps_1 A\otimes Z}
            \paren*{e^{-it\widetilde H}\otimes I}.
    \]
    The first operation is known. The second is, up to a global phase, forward evolution under \(H\) for time \(\frac{2\eps t}{\alpha}\). Therefore, a tester using total forward evolution time \(T\) distinguishes zero from the random ensemble using \(\frac{2\eps}{\alpha}T\) total forward evolution time. By~\cite[Theorem 3.10]{GLZ26},
    \[
        \frac{2\eps}{\alpha}T=\bOm{\frac{1}{\alpha^2}},
    \]
    and hence
    \[
        T=\bOm{\frac{1}{\alpha\eps}}
        =\bOm{\min\set*{\frac{1}{\eps^2},\frac{2^{\frac{n}{2}}}{\eps}}}.
    \]
    Since \(\eps_2(\eps_2-\eps_1) \le \eps^2 \le 2\eps_2(\eps_2-\eps_1)\), this gives the claimed bound.

    The bound for Hamiltonian certification follows by setting \(s=0\) in the sparsity bound and \(H_0=0\).
\end{proof}

\section{Acknowledgements}
This project is supported by the National Research Foundation, Singapore
through the National Quantum Office, hosted in A*STAR, under its Advanced
Quantum Algorithms and Solutions Funding Initiative (S25Q9DA001 and
S25Q9DA002).

\subsection{AI}
The key technical result in this paper, the ability to Pauli sample with
minimal overhead, was discovered, almost by accident, while using ChatGPT Astra
to check details of a substantially more complicated algorithm we had developed
for sparsity testing. The implications of this fact were worked out by the
authors, with assistance from Astra. Astra was also used to check the
manuscript for errors. The text of the manuscript, including the proofs, was
written by the authors, who take full responsibility for the content.

\newpage
\bibliographystyle{alpha}
\bibliography{refs}

\newcommand{\etalchar}[1]{$^{#1}$}
\begin{thebibliography}{GJW{\etalchar{+}}26}

\bibitem[ACC{\etalchar{+}}26]{allen2026power}
Richard~R. Allen, Matthias~C. Caro, Yanlin Chen, Tom Gur, and Angus Lowe.
\newblock The power of time reversal in {H}amiltonian property testing, 2026.

\bibitem[ADEG25]{AAEG25}
Srinivasan Arunachalam, Arkopal Dutt, and Francisco Escudero~Guti{\'e}rrez.
\newblock Testing and learning structured quantum {H}amiltonians.
\newblock In {\em Proceedings of the 57th Annual ACM Symposium on Theory of Computing}, pages 1263--1270, 2025.

\bibitem[BCO26]{BCO24}
Andreas Bluhm, Matthias~C. Caro, and Aadil Oufkir.
\newblock Hamiltonian property testing.
\newblock {\em Quantum}, 10:1979, January 2026.

\bibitem[BLY{\etalchar{+}}25]{BLLYZ25}
Zongbo Bao, Yuxuan Liu, Penghui Yao, Zekun Ye, and Jialin Zhang.
\newblock Efficient non-adaptive quantum algorithms for tolerant junta testing.
\newblock {\em arXiv preprint arXiv:2508.17306}, 2025.

\bibitem[Che52]{chernoff1952measure}
Herman Chernoff.
\newblock A measure of asymptotic efficiency for tests of a hypothesis based on the sum of observations.
\newblock {\em The Annals of Mathematical Statistics}, 23(4):493--507, 1952.

\bibitem[Che26]{C26}
Kenny Chen.
\newblock Exponential quantum advantage in testing {F}ourier dimensionality.
\newblock {\em arXiv preprint arXiv:2609.25816}, 2026.

\bibitem[Cho75]{Choi75}
Man-Duen Choi.
\newblock Completely positive linear maps on complex matrices.
\newblock {\em Linear Algebra and its Applications}, 10(3):285--290, 1975.

\bibitem[CY26]{CY26}
Zhili Chen and Zhan Yu.
\newblock Learning arbitrary {L}indbladians from time evolution.
\newblock {\em arXiv preprint arXiv:2607.28610}, 2026.

\bibitem[DB08]{dahlquist2008numerical}
Germund Dahlquist and {\AA}ke Bj{\"o}rck.
\newblock {\em Numerical Methods in Scientific Computing}, volume~1.
\newblock Society for Industrial and Applied Mathematics, 2008.

\bibitem[FLT23]{Fang_2023}
Di~Fang, Lin Lin, and Yu~Tong.
\newblock Time-marching based quantum solvers for time-dependent linear differential equations.
\newblock {\em Quantum}, 7:955, March 2023.

\bibitem[GJW{\etalchar{+}}26]{GJWYZ26}
Minbo Gao, Zhengfeng Ji, Qisheng Wang, Wenjun Yu, and Qi~Zhao.
\newblock Quantum {Hamiltonian} certification.
\newblock In {\em Proceedings of the 2026 Annual ACM-SIAM Symposium on Discrete Algorithms (SODA)}, pages 1424--1467. SIAM, 2026.

\bibitem[GLZ26]{GLZ26}
Francisco~Escudero Gutiérrez, Junseo Lee, and Sebastian Zur.
\newblock Hamiltonian locality testing and certification do not achieve the {H}eisenberg limit.
\newblock {\em arXiv preprint arXiv:2610.03205}, 2026.

\bibitem[GSLW19]{GSLW19}
András Gilyén, Yuan Su, Guang~Hao Low, and Nathan Wiebe.
\newblock Quantum singular value transformation and beyond: exponential improvements for quantum matrix arithmetics.
\newblock In {\em Proceedings of the 51st Annual ACM SIGACT Symposium on Theory of Computing}, STOC ’19, page 193–204. ACM, June 2019.
\newblock \href{https://arxiv.org/abs/1806.01838}{arXiv:1806.01838}.

\bibitem[Hoe63]{hoeffding1963probability}
Wassily Hoeffding.
\newblock Probability inequalities for sums of bounded random variables.
\newblock {\em Journal of the American Statistical Association}, 58(301):13--30, 1963.

\bibitem[Jam72]{Jam72}
A.~Jamiołkowski.
\newblock Linear transformations which preserve trace and positive semidefiniteness of operators.
\newblock {\em Reports on Mathematical Physics}, 3(4):275--278, 1972.

\bibitem[KL25]{KL25}
John Kallaugher and Daniel Liang.
\newblock {Hamiltonian Locality Testing via Trotterized Postselection}.
\newblock In Bill Fefferman, editor, {\em 20th Conference on the Theory of Quantum Computation, Communication and Cryptography (TQC 2025)}, volume 350 of {\em Leibniz International Proceedings in Informatics (LIPIcs)}, pages 10:1--10:20, Dagstuhl, Germany, 2025. Schloss Dagstuhl -- Leibniz-Zentrum f{\"u}r Informatik.

\bibitem[MdW16]{MdW16}
Ashley Montanaro and Ronald de~Wolf.
\newblock A survey of quantum property testing.
\newblock {\em Theory of Computing Graduate Surveys}, 7:1--81, 2016.

\bibitem[ST25]{ST25}
Savar~D. Sinha and Yu~Tong.
\newblock Improved {Hamiltonian} learning and sparsity testing through {Bell} sampling.
\newblock {\em arXiv preprint arXiv:2509.07937}, 2025.

\bibitem[Tre19]{trefethen2019approximation}
Lloyd~N. Trefethen.
\newblock {\em Approximation Theory and Approximation Practice}.
\newblock Society for Industrial and Applied Mathematics, extended edition, 2019.

\bibitem[TW25]{TW25}
Ewin Tang and John Wright.
\newblock Amplitude amplification and estimation require inverses.
\newblock {\em arXiv preprint arXiv:2507.23787}, 2025.

\bibitem[Zha25]{Z25}
Andrew Zhao.
\newblock Learning the structure of any {H}amiltonian from minimal assumptions.
\newblock In {\em Proceedings of the 57th Annual ACM Symposium on Theory of Computing}, STOC ’25, page 1201–1211. ACM, June 2025.
\newblock \href{https://arxiv.org/abs/2410.21635}{arXiv:2410.21635}.

\end{thebibliography}

\newpage
\appendix
\section{Hamiltonian Norm Estimation}
\label{app:hamiltonian-norm-estimation}
In this section, we describe an algorithm that either detects that \(\norm{H}_2^2 \le \eps_2^2\), or estimates \(\norm{H}_2^2\) to multiplicative error \(\bO{\frac{\eps_2^2 - \eps_1^2}{\eps_2^2}}\). First, we need the following lemma.

\begin{lemma}\label{lem:cosine-series}
    For all \(x \in [-2, 2]\), \(\frac{x^2}{2}=\sum_{j=1}^{\infty}w_j(1-\cos(jx))\) where \(w_j=\frac{8(-1)^{j+1}\sin^2\left(\frac{j}{2}\right)}{j^4}.\)
\end{lemma}

\begin{proof}
    Consider the cosine series
    \[
        \frac{y^2}{2}
        =\frac{\pi^2}{6}
        +2\sum_{j=1}^{\infty}\frac{(-1)^j}{j^2}\cos(jy),
        \qquad \abs{y}\le\pi.
    \]
    The constant term is the average of \(\frac{y^2}{2}\) on \([-\pi,\pi]\), and the coefficient of \(\cos(jy)\) is
    \[
        \frac{1}{\pi}\int_{-\pi}^{\pi}\frac{y^2}{2}\cos(jy)\,dy
        =\frac{2(-1)^j}{j^2}.
    \]
    The series is absolutely convergent, so we may integrate it term by term. Now, observe that, for any \(x\in[-2,2]\) and \(u\in[0,1]\),
    \[
        \frac{1}{2}\left(
        \frac{(x+u)^2}{2}+\frac{(x-u)^2}{2}
        \right)-\frac{u^2}{2}
        =\frac{x^2}{2}.
    \]
    Moreover, applying the same operation to the cosine series---taking the average of its values at \(x+u\) and \(x-u\), and then subtracting its value at \(u\)---gives
    \[
        \frac{\cos(j(x+u))+\cos(j(x-u))}{2}-\cos(ju)
        =(\cos(jx)-1)\cos(ju).
    \]
    All three arguments lie in \([-\pi,\pi]\), since \(\abs{x\pm u}\le3<\pi\). Therefore, for every \(u\in[0,1]\),
    \[
        \frac{x^2}{2}
        =2\sum_{j=1}^{\infty}\frac{(-1)^j}{j^2}
        (\cos(jx)-1)\cos(ju).
    \]
    Now multiply both sides by \(2(1-u)\) and integrate from \(0\) to \(1\). The left-hand side remains \(\frac{x^2}{2}\). On the right-hand side, each factor \(\cos(ju)\) is replaced by
    \[
        2\int_0^1(1-u)\cos(ju)\,du
        =\frac{2(1-\cos j)}{j^2}
        =\frac{4\sin^2\left(\frac{j}{2}\right)}{j^2}.
    \]
    Combining the two calculations gives us
    \begin{align*}
        \frac{x^2}{2}
        &=2\sum_{j=1}^{\infty}\frac{(-1)^j}{j^2}
        (\cos(jx)-1)
        \frac{4\sin^2\left(\frac{j}{2}\right)}{j^2}\\
        &=\sum_{j=1}^{\infty}w_j(1-\cos(jx)),
        \qquad
        \text{where}~w_j=\frac{8(-1)^{j+1}\sin^2\left(\frac{j}{2}\right)}{j^4}.
    \end{align*}
\end{proof}

We are now ready to describe our estimator.

\begin{lemma}\label{lem:norm-estimate}
    For all \(0 \le \eps_1 < \eps_2 < 1\) and \(\delta \in (0,1/2)\), there is an algorithm that either certifies that \(\norm{H}_2^2 < \eps_2^2\), or outputs a \(\left(1 \pm \frac{\eps_2^2 - \eps_1^2}{100\eps_2^2}\right)\)-approximation of \(\norm{H}_2^2\), with probability at least \(1 - \delta\), using
    \[
        \bO{\frac{1}{(\eps_2 - \eps_1)^2}\log{\left(\frac{1}{\delta}\right)}}
    \]
    total Hamiltonian evolution time and queries to the time evolution operator.
\end{lemma}

\begin{proof}
    We estimate Bell-sampling probabilities at several fixed times and take a weighted sum of the estimates.

    Let \(q(t)\) be the probability of observing a non-identity Pauli when sampling from \(e^{-itH}\) using~\cref{lem:bell-sampling}. If \(\lambda_1,\ldots,\lambda_{2^n}\) are the eigenvalues of \(H\), then, since \(\Tr(e^{-itH})=\sum_{r=1}^{2^n}e^{-it\lambda_r}\), we have
    \begin{align*}
        q(t)
        &= 1-\abs*{\frac{\Tr(e^{-itH})}{2^n}}^2 \\
        &= \frac{1}{2^{2n}}\sum_{r,s=1}^{2^n}
        \left(1-e^{-it(\lambda_r-\lambda_s)}\right) \\
        &= \frac{1}{2^{2n}}\sum_{r,s=1}^{2^n}
        \left(1-\cos\left(t(\lambda_r-\lambda_s)\right)\right).
    \end{align*}
    
    To recover the norm exactly, we express \(\frac{x^2}{2}\) as a sum of \(1-\cos(jx)\) for \(\abs{x}\le2\), using~\cref{lem:cosine-series}. By~\cref{lem:cosine-series}, for any \(x \in [-2, 2]\),
    \[
        \frac{x^2}{2}=\sum_{j=1}^{\infty}w_j(1-\cos(jx)).
    \]
    where \(w_j=\frac{8(-1)^{j+1}\sin^2\left(\frac{j}{2}\right)}{j^4}.\) Each eigenvalue difference satisfies \(\abs{\lambda_r-\lambda_s}\le2\). Substituting \(x=\lambda_r-\lambda_s\) in the preceding equation gives
    \[
        \sum_{j=1}^{\infty}w_j
        \left(1-\cos\left(j(\lambda_r-\lambda_s)\right)\right)
        =\frac{(\lambda_r-\lambda_s)^2}{2}.
    \]
    Averaging this identity over \(r,s\), and interchanging the finite sum over \(r,s\) with the convergent series over \(j\), we obtain
    \begin{align*}
        \sum_{j=1}^{\infty}w_jq(j)
        &= \frac{1}{2^{2n}}\sum_{r,s=1}^{2^n}
        \sum_{j=1}^{\infty}w_j
        \left(1-\cos\left(j(\lambda_r-\lambda_s)\right)\right) \\
        &= \frac{1}{2^{2n}}\sum_{r,s=1}^{2^n}
        \frac{(\lambda_r-\lambda_s)^2}{2} \\
        &= \frac{1}{2^n}\sum_{r=1}^{2^n}\lambda_r^2
        -\left(\frac{1}{2^n}\sum_{r=1}^{2^n}\lambda_r\right)^2 \\
        &= \frac{\Tr(H^2)}{2^n}
        -\left(\frac{\Tr(H)}{2^n}\right)^2 \\
        &= \norm{H}_2^2,
    \end{align*}
    where the last equality uses \(\Tr(H)=0\) and the definition of the normalized Schatten \(2\)-norm.

    We use only the first
    \[
        K=\ceil*{\frac{10^4\eps_2^2}{\eps_2^2-\eps_1^2}}
    \]
    times. The other terms contribute at most
    \[
        \abs*{\norm{H}_2^2-\sum_{j=1}^K w_jq(j)}
        \le8\norm{H}_2^2\sum_{j>K}\frac{1}{j^2}
        \le\frac{8\norm{H}_2^2}{K}.
    \]
    We allow larger estimation errors at later times because their weights decrease as \(j^{-4}\). For each \(1\le j\le K\), take
    \[
        N_j=\ceil*{
        \frac{10^9\eps_2^2}{j^3(\eps_2^2-\eps_1^2)^2}
        \log{\left(\frac{4j^2}{\delta}\right)}}
    \]
    independent samples from \(e^{-ijH}\) using~\cref{lem:bell-sampling}, and let \(\hat q_j\) be the fraction of non-identity outcomes. By~\cref{lem:bernstein-prelim}, with failure probability at most \(\frac{\delta}{2j^2}\),
    \[
        \abs{\hat q_j-q(j)}
        \le
        \frac{j^{\frac{5}{2}}(\eps_2^2-\eps_1^2)}{10^4\eps_2^2}
        \max\{\norm{H}_2^2,\eps_2^2\}.
    \]
    A union bound gives all these estimates simultaneously with probability at least \(1-\delta\).

    Let \(z=\sum_{j=1}^K w_j\hat q_j\). We have that
    \begin{align*}
        \abs{z-\norm{H}_2^2}
        &\le\frac{8\norm{H}_2^2}{K}
        +\frac{8(\eps_2^2-\eps_1^2)}{10^4\eps_2^2}
        \max\{\norm{H}_2^2,\eps_2^2\}
        \sum_{j=1}^K\frac{1}{j^{\frac{3}{2}}}\\
        &\le\frac{\eps_2^2-\eps_1^2}{200\eps_2^2}
        \max\{\norm{H}_2^2,\eps_2^2\},
    \end{align*}
    where \(\sum_{j\ge1}\frac{1}{j^{\frac{3}{2}}}
    \le1+\int_1^\infty x^{-\frac{3}{2}}\,dx=3\).

    If \(z\le\frac{3\eps_2^2}{4}\), then certify \(\norm{H}_2^2<\eps_2^2\); otherwise, return \(z\). Therefore, when an estimate is returned, \(\norm{H}_2^2>\frac{\eps_2^2}{2}\), and its error is at most
    \[
        \frac{\eps_2^2-\eps_1^2}{100\eps_2^2}\norm{H}_2^2.
    \]
    Finally, the total evolution time is
    \[
        \sum_{j=1}^K jN_j
        \le\frac{K(K+1)}{2}
        +\frac{10^9\eps_2^2}{(\eps_2^2-\eps_1^2)^2}
        \sum_{j=1}^K\frac{\log{\left(\frac{4j^2}{\delta}\right)}}{j^2}.
    \]
    Both \(\sum_{j\ge1}\frac{1}{j^2}\) and \(\sum_{j\ge1}\frac{\log j}{j^2}\) converge, and
    \(K^2=\bO{\frac{\eps_2^4}{(\eps_2^2-\eps_1^2)^2}}\).
    Since \(\eps_2<1\), this gives total evolution time
    \[
        \bO{\frac{\eps_2^2}{(\eps_2^2-\eps_1^2)^2}
        \log{\left(\frac{1}{\delta}\right)}}
        =
        \bO{\frac{1}{(\eps_2-\eps_1)^2}
        \log{\left(\frac{1}{\delta}\right)}}.
    \]
    The number of queries is no greater, since each application of the time evolution operator uses at least constant time.
\end{proof}

\end{document}